\documentclass[12pt]{iopart}

\usepackage[nomarkers,nolists]{endfloat} 
 
\usepackage[dvips]{epsfig}
\usepackage{graphicx} 
\usepackage{cite}
\usepackage{amssymb}
\usepackage[usenames]{color}
\begin{document}

\title{Geometric Phase Effects in the Ultracold D + HD $\to$ D + HD and D + HD $\leftrightarrow$ H + D$_2$  Reactions}
\author{B. K. Kendrick}
\address{Theoretical Division (T-1, MS B221), Los Alamos National Laboratory, Los Alamos,
New Mexico 87545, USA}
\author{Jisha Hazra}
\address{Department of Chemistry, University of Nevada, Las Vegas, Nevada 89154, USA}
\author{N. Balakrishnan} 
\address{Department of Chemistry, University of Nevada, Las Vegas, Nevada 89154, USA}
\vskip 10pt
\begin{abstract}
The results of accurate quantum reactive scattering calculations for the 
D + HD($v=4$, $j=0$) $\to$ D + HD($v'$, $j'$), 
D + HD($v=4$, $j=0$) $\to$ H + D$_2$($v'$, $j'$)
and H + D$_2$($v=4$, $j=0$) $\to$ D + HD($v'$,$j'$) reactions 
are presented for collision energies between $1\,\mu{\rm K}$ and $100\,{\rm K}$.
The {\it ab initio} BKMP2 PES for the ground electronic state of H$_3$ is used and
all values of total angular momentum between $J=0-4$ are included.
The general vector potential approach is used to include the geometric phase.
The rotationally resolved, vibrationally resolved, and total reaction rate coefficients 
are reported as a function of collision energy.
Rotationally resolved differential cross sections are also reported as a function of collision energy and scattering angle.
Large geometric phase effects appear in the ultracold reaction rate coefficients which result 
in a significant enhancement or suppression of the rate coefficient  (up to $3$ orders of magnitude) 
relative to calculations which ignore the geometric phase.
The results are interpreted using a new quantum interference mechanism which is
unique to ultracold collisions.
Significant effects of the geometric phase also appear in the rotationally resolved differential cross sections
which lead to a very different oscillatory structure in both energy and scattering angle.
Several shape resonances occur in the $1$ - $10\,{\rm K}$ energy range and
the geometric phase is shown to significantly alter the predicted resonance spectrum.
The geometric phase effects depend sensitively on the nuclear spin which may provide experimentalists with
the ability to control the reaction by the selection of a particular nuclear spin state.
\end{abstract}

\maketitle

\section{Introduction}
\label{sect1}

For over 89 years, the Born-Oppenheimer\cite{BornOppen} method has been the foundation for 
the quantum mechanical treatment of molecular structure, spectra, and scattering.
This methodology is based on a power series expansion of the molecular wave function and
energies in terms of the small electron to nuclei mass ratio ${\kappa} = (m_e/m_{\rm nuc})^{1/4}$.
The lowest order terms give the well known two-step approach for solving the
quantum mechanical molecular problem.  In the first step, the electronic Schr\"odinger equation
is solved for a given fixed position of the nuclei.  Repeated solutions of the electronic structure 
are performed on a grid of nuclear geometries to construct an effective electronic potential
energy surface (PES).  This discrete set of points is usually fit to appropriate analytic functions 
to produce a smooth surface which can then be evaluated for any nuclear geometry.
In the second step, the effective Schr\"odinger equation for the nuclear motion is solved 
using the PES computed in the first step (which is typically for the ground electronic state).
In the absence of electronic degeneracies and when the couplings to excited electronic states
can be ignored, this approach works very well as documented by the large body of 
impressive comparisons between theory and experiment.\cite{HerzbergI,HerzbergII}
In the presence of electronic degeneracies or when the couplings
to excited electronic states become important, then the approach discussed above must
be generalized to include these excited electronic states and their couplings.
For example, if the ground electronic state becomes degenerate with 
an excited electronic state for some nuclear geometry (i.e., it exhibits a conical 
intersection), then the real-valued ground state electronic wave function changes sign for any nuclear 
motion which encircles the degeneracy.\cite{LH58,herzberg63}
This sign change occurs even though the degeneracy itself 
may lie very high in energy and is not energetically accessible for the given kinetic energy of the nuclear motion.
Only the minimum energy pathway which encircles the degeneracy need be energetically accessible.
Mead and Truhlar\cite{meadtruhlar79} showed that in this case the effects of the electronic sign change
(i.e., double-valuedness) on the nuclear motion can be included by transforming to
a {\it complex} single-valued electronic basis which gives rise to a generalized
momentum operator in the Schr\"odinger equation for the nuclear motion.
That is, the nuclear motion momentum operator ${\bf p}$ becomes ${\bf p} - {\bf A}$ where ${\bf A}$ is an effective $U(1)$ 
vector (or gauge) potential analogous to that of a magnetic field.
The vector potential ${\bf A}$ is not associated with a real magnetic field but it has
the same mathematical properties.  In particular, the mathematical form of ${\bf A}$ is equivalent to that 
of a magnetic solenoid centered at the point of degeneracy.
Thus, the nuclear motion is governed not only by an effective electronic PES but also
by the presence of an effective magnetic field ${\bf B} = \nabla \times {\bf A}$.
This magnetic field has the peculiar property of being zero everywhere except
at the point of degeneracy where it exhibits a delta function singularity.
Integrating the vector potential along a closed path which encircles the degeneracy
gives the phase shift: ${\rm exp}[i\,\oint_{\cal C} {\bf A} \cdot {\bf dl}] = {\rm exp}[i\,\int_S {\bf B}\cdot{\bf dS}] = {\rm exp}[i\pi]$ (i.e. $-1$).
That is, the phase shift is equal to the flux of the magnetic field through the surface $S$ enclosed by the path $\cal C$.
Due to its geometrical origin (i.e., as the holonomy\cite{bsimon84,bohm92} associated with a non-trivial curvature two-form or gauge field), 
this phase shift is often referred to as the geometric phase (GP).\cite{meadRMP,arno}
Mead recognized the similarity between the molecular vector potential and its associated GP
with the Aharonov-Bohm effect.\cite{ABeffect}
Thus, he originally referred to it as the ``Molecular Aharonov-Bohm'' (MAB) effect.\cite{meadBS_MAB}
Berry\cite{berry84} later generalized the molecular treatment to a general quantum system undergoing adiabatic
time evolution. He also considered other example systems for which different kinds of vector potentials appear (i.e., monopoles).
Hence, the GP is also often referred to as Berry's phase.

Prior to the work of Mead and Truhlar, the effects of the electronic sign change (or GP) were included in 
molecular spectra by using a real double-valued nuclear motion basis set within a two-electronic state model.\cite{LH58,obrien64}
However, a two-state approach requires twice the number of nuclear motion basis functions which typically results in $2^3=8$ times more computational work.
In molecular systems for which the couplings to the excited electronic states can be ignored, this additional computational expense is unnecessary and
for heavy nuclei and systems with deep attractive wells the added computational expense can be prohibitive.
In contrast, the vector potential approach includes the GP using a single (ground state) electronic PES. 
Furthermore, the vector potential approach provides a consistent and transparent treatment of identical particle permutation symmetry 
(which was the original motivation for pursing this approach).\cite{meadtruhlar79,meadRMP,kendrick1995}
Mead showed that the vector potential approach gives the correct molecular spectra, namely the ground vibrational state for X$_3$ systems is of E symmetry instead of A$_1$ or A$_2$.\cite{meadBS_MAB}
 He also demonstrated gauge invariance of the computed spectra with respect to $U(1)$ gauge transformations.\cite{meadBS_MAB}
For molecular systems with special symmetry (i.e., H$_3$ with three identical nuclei and a conical intersection located at the equilateral (D$_{3h}$) geometry),
the GP also can be included for a single (ground state) electronic PES by using a real double-valued nuclear motion basis set instead of the vector potential approach.\cite{obrien64,meadtruhlar79,kuppermannJ01990}
However, in more general situations where the conical intersection is not located at a symmetry point or there are multiple intersections (as in HO$_2$),
then the vector potential approach is advantageous.\cite{meadtruhlar79,meadRMP,kendrick1996I,kendrick1996II,kendrickFeature2003}
In molecular systems for which the couplings to the excited electronic state are important, then a fully coupled 2x2 treatment is required with its associated computational expense. 
In this case, the derivative couplings between the two (or in general $N$) electronic states can be expressed as a non-abelian $U(N)$ gauge potential (i.e., a non-commuting matrix operator).
Mead and Truhlar showed that in general the truncation to an incomplete $N$-dimensional electronic subspace results in a non-trivial gauge field given by $F^{ij}_{nm}=\partial_i  A^j_{nm} - \partial_j A^i_{nm} - i\,[A^i_{nl},A^j_{lm}]$ where the third term is a commutator.\cite{meadtruhlar82}
They also showed that for a complete electronic space the non-abelian gauge potential ${\bf A}_{nm}$ is a pure gauge for which $F^{ij}_{nm}=0$.\cite{meadtruhlar82}
Thus, the truncation to a finite dimensional (incomplete) electronic subspace induces a non-zero curvature or gauge field.
In this paper, we focus on the GP arising from an electronic degeneracy (conical intersection) and consider only the $U(1)$ gauge potential.
For more details on the non-abelian $U(N)$ case, we refer the interested reader to several treatments in the literature.\cite{meadRMP,kendrickFeature2003,meadtruhlar82,mead83,mead87,zygelman1987,zygelman1990,BmByKendrick1992,kendrick2002,zygelman2015}

GP effects in molecular spectra have been reported in several theoretical treatments\cite{LH58,meadBS_MAB,gerber1978,thompson1985,truhlar1986,ham1987,koppel1990,mayer1996,schon1994,schon1995,kendrick1997,kendrick1997B,babikov2005}
and it has been confirmed experimentally for several molecules Cu$_3$,\cite{valentini1986} 
Li$_3$,\cite{demtroder1998} and Na$_3$.\cite{demtroder2000}
In stark contrast, the experimental measurement of a GP effect in molecular scattering has continued to be elusive to this day.\cite{zare2013,zare2013B}
The first theoretical prediction of a GP effect in scattering was made by Mead for the H + H$_2$ system.\cite{meadScatter}
He showed that the GP changes the sign on the interference term between the non-reactive and reactive contributions to the scattering amplitude
for the para-para and ortho-ortho transitions.
The sign change alters the oscillatory pattern of the theoretically computed differential cross sections (DCSs).
However, due to the experimental difficulties associated with this system, experimental confirmation of Mead's prediction has not been made.
Initial theoretical\cite{WuKuppermann1993,WuKuppermann1995A,WuKuppermann1995B} 
studies reported significant GP effects in both the integral and DCSs for the isotopic reactions H + D$_2$ $\to$ D + HD and D + H$_2$ $\to$ H + HD.
The geometric phase effects were claimed to resolve the reported discrepancies between theory and experiment.\cite{zare1991,zare1992,zare1993}
However, Kendrick later showed that the GP effects initially reported for the isotopic reactions all but cancel out in 
both the integral and DCSs when the partial cross sections are summed over the total angular momentum $J$ to obtain fully converged cross sections.\cite{kendrickHD2GPJ5,kendrickHD2GPJ,kendrickDH2GPJ,kendrickFeature2003}
In addition, excellent agreement was reported between the theoretically computed integral and DCSs computed {\it without} the GP 
and high resolution crossed molecular beam experiments.\cite{wrede1997,miranda1998,kendrickHD2GPJ}
The cancellation of the GP effect with respect to the partial wave sum was confirmed by Althorpe and co-workers using an entirely different computational methodology.\cite{carlosAlthorpe2005,Althorpescience2005,Althorpe2006,Althorpe2008,Boukline2014}
At higher scattering energies but below the energy of the conical intersection, Althorpe and coworkers found small oscillations in the DCSs due to the GP.\cite{carlosAlthorpe2005,BoulkinH3,bouaklineGP_2008}
At energies above the conical intersection, large GP effects on the DCSs were reported which give rise to broader bimodal features.\cite{BoulkinH3,bouaklineGP_2008,bouaklineQCT_2010}
Unfortunately, a recent experimental effort was unable to resolve the GP oscillations in the DCSs for the H + HD $\to$ H + HD reaction at energies below the conical intersection.\cite{zare2013,zare2013B}
No significant GP effects have been reported in the integral cross sections (or reaction rate coefficients) at any thermal energy.

Until recently,\cite{Kendrik15_1,Kendrik15_2,hazra2015JPC,hazra2016JPhysA} all of the previous studies of GP effects in molecular scattering were done at thermal energies.
At cold and ultracold collision energies (which correspond to temperatures below $1\,{\rm K}$ ($8.6\times 10^{-5}\,{\rm eV}$) and $1\,{\rm mK}$ ($8.6\times 10^{-8}\,{\rm eV}$), respectively)
the collision outcome is governed by enhanced quantum mechanical effects which include tunneling, resonances, symmetry, and interference.
Experimental capabilities for cooling and trapping of molecules
have made rapid progress in recent years enabling the exploration of this entirely new and exciting energy regime.\cite{coldMolBook,carr09,ospelkaus2010,knoop2010,bala2016}
In the ultracold regime, a single partial wave (i.e., the $l=0$ angular momentum partial wave or $s$-wave for bosons and distinguishable particles, and $l=1$ or $p$-wave for identical fermions) contributes to the scattering cross sections and
the ultracold reaction rate coefficients obey the well known Bethe-Wigner threshold laws.\cite{Bethe35,Wigner48,schwenke1985,stwalley2004,weck06}
For exoergic processes these rate coefficients approach finite measurable values comparable to or even larger than their values at thermal energies.\cite{Kendrik15_1,Kendrik15_2,hazra2015JPC,hazra2016JPhysA,Quemener08,Quemener09,Juan11,Pradhan13,LiYb2015}
The tiny kinetic energy associated with ultracold collisions makes them amenable to control via external electric or 
magnetic fields.\cite{coldMolBook,carr09,ospelkaus2010,knoop2010,Krems2005,Tscherbul2008,Tscherbul2015,quemener12}
In contrast to collisions at thermal energies, the unique properties associated with ultracold collisions can also lead to dramatic GP effects in 
{\it both} the integral and DCSs, as reported in our recent work on the ultracold barrierless reactions:  
O + OH($v=0$, $j=0$) $\to$ H + O$_2$($v'$,$j'$), H + H$_2$($v=4$, $j=0$) $\to$ H + H$_2$($v'$,$j'$),
H/D + HD($v=4$,$j=0$) $\to$ H/D + HD($v'$,$j'$), H + HD($v=4$,$j=0$) $\to$ H + D$_2$($v'$,$j'$)
and H + D$_2$($v=4$,$j=0$) $\to$ H + HD($v'$,$j'$).\cite{Kendrik15_1,Kendrik15_2,hazra2015JPC,hazra2016JPhysA}
Vibrational excitation of the reactants in the H$_3$ system leads to a barrierless reaction pathway along the vibrational adiabats 
which proceeds over a potential well.\cite{truhlar1970,miller1980,truhlar1996,jankunas_PNAS2014}
In particular, the dynamics changes from a barrier reaction to a barrierless one for $v>3$ which can lead to significant reactivity at ultracold collision energies.\cite{Simbotin2011,Simbotin2014,Simbotin2015}
Two properties in particular contribute to enhanced GP effects in the ultracold regime:
(1) Isotropic ($s$-wave) scattering for which the scattering occurs at all angles and can lead to maximum constructive or destructive interference,
and (2) The relative phase between the two interfering scattering amplitudes which encircle the conical intersection preferentially approaches an integral multiple of $\pi$.
If the magnitudes of the two interfering scattering amplitudes are comparable, then maximum constructive or destructive 
interference can occur depending upon whether the relative phase approaches an even or odd multiple of $\pi$, respectively.
Since the GP alters the sign of the interference term or equivalently shifts the relative phase by $\pi$, the opposite interference occurs when the GP is included
(i.e, the interference becomes constructive instead of destructive and vice versa). Thus, the GP acts like a quantum switch turning the reaction on or off.\cite{Kendrik15_1}

In this paper we report GP effects in the cold/ultracold 
D + HD($v=4$,$j=0$) $\to$ D + HD($v'$,$j'$), D + HD($v=4$,$j=0$) $\to$ H + D$_2$($v'$, $j'$)
and H + D$_2$($v=4$, $j=0$) $\to$ D + HD($v'$,$j'$) reactions for collision energies between $1\,\mu{\rm K}$ ($8.6\times 10^{-11}\,{\rm eV}$) and $100\,{\rm K}$ ($8.6\times 10^{-3}\,{\rm eV}$).
As in prior work, the vector potential approach is used to include the GP.\cite{meadtruhlar79,kendrick1996I}
The previous calculations\cite{Kendrik15_2} have been extended to include all values of total angular momentum $J=0-4$. 
GP effects on the DCSs are reported for the first time as a function of collision energy and scattering angle.
Total as well as rotationally and vibrationally resolved reaction rate coefficients are also reported as a function of collision energy for many product states.
It is shown that the large GP effect (over three orders of magnitude in some cases) effectively controls the outcome of the ultracold hydrogen exchange reactions.
In addition, shape resonances are predicted to occur at higher collision energies between $1\,{\rm K}$ and $10\,{\rm K}$.
Low energy shape resonances have been previously reported for D + H$_2$,\cite{Simbotin2015} and have also been experimentally measured in
inelastic collisions of O$_2$-H$_2$\cite{chefdeville2013} and NO-He.\cite{vogels2015}
It is shown that the GP significantly alters the predicted resonance spectrum for the hydrogen exchange reactions.
As mentioned above, the origin of the large GP effect is discussed in terms of a newly discovered quantum interference mechanism which is unique to ultracold collisions.\cite{Kendrik15_1}
The enhancement or suppression of the reaction rate is shown to depend upon the nuclear spin. Thus, experimentalist might control the outcome by the selection of a particular nuclear spin state.
The paper is organized as follows:  the computational methodology is discussed in Section \ref{sect2} followed by the scattering results for each system in Section \ref{sect3}.
Sections \ref{sect3a}, \ref{sect3b}, and \ref{sect3c} present results and discussion for the D+HD $\to$ D+HD, D+HD $\to$ H + D$_2$, and 
H + D$_2$ $\to$ D + HD reactions, respectively.
The conclusions are discussed in Sect. \ref{sect4}.

\section{Computational Method}
\label{sect2}
The calculations were performed using the LANL APH3D quantum reactive scattering code 
which is a time-independent coupled-channel formalism based on the Adiabatically adjusting 
Principal axis Hyperspherical (APH) approach of Pack and Parker.\cite{Pack87}
For computational efficiency, Smith-Johnson\cite{kendrick99,smith1962,johnson1980} hyperspherical coordinates are used 
in the three-body interaction region and Delves\cite{parker2002,Delves1959,Smith1960} hyperpsherical coordinates are used 
in the long-range region where the diatomic channels are well separated (non-interacting).
Specialized body-frame basis functions are used for accurately treating non-zero total
angular momentum ($J$) and the associated Eckart singularities.\cite{kendrick99}
The geometric phase is also included using the general vector potential approach of
Mead and Truhlar.\cite{meadtruhlar79,kendrick1996I,kendrick1996II,kendrickHD2GPJ5}
The methodology is numerically exact for a given Born-Oppenheimer PES (i.e., no 
dynamical approximations are used) and the computer code has been parallelized to run efficiently on a variety
of computational platforms from a single workstation to massively parallel 
supercomputers.\cite{kendrickHD2GPJ5,kendrickHD2GPJ,kendrickDH2GPJ,kendrick99}
The methodology can also be used to compute molecular spectra\cite{kendrick1997,kendrick1997B,babikov2005,kendrickHO297}
and is well suited for treating ultracold reactions.\cite{Kendrik15_1,Kendrik15_2,hazra2015JPC,hazra2016JPhysA,Quemener08,Quemener09,Juan11,Pradhan13,LiYb2015}

The explicit expressions for the hyperspherical coupled-channel equations and detailed discussion of their numerical solution
are discussed in prior work.\cite{Pack87,kendrick1996I,kendrick99,kendrickHD2GPJ5,kendrickHD2GPJ,kendrickDH2GPJ}
Thus, we give only a brief summary of the methodology here.
The six-dimensional (6D) quantum three-body problem is solved numerically by first discretizing the hyperradius $\rho$ 
and solving the five-dimensional (5D) angular (surface function) eigenvalue problem
at each fixed value of $\rho_\xi$.  The 5D angular solutions provide the basis set for the
coupled-channel solutions and are accurate in a small region (sector) centered about each $\rho_\xi$.
The potential coupling matrices within each sector and the overlap matrices between the 
surface functions at adjacent sectors are computed using the surface functions at each $\rho_\xi$.
The 5D surface function eigenvalue problem is solved using the efficient sparse matrix 
diagonalization routine PARPACK with Chebychev preconditioning.\cite{kendrick99,kendrickHD2GPJ5,Pendergast1994}
PARPACK is a parallel implementation of the implicitly restarted Lanczos method (IRLM).\cite{sorensen1992,sorensenARPACK,sorensenPARA96,sorensenUSER}
In this approach, the multiplication of the Hamiltonian matrix
on a vector is all that is required (i.e., the 5D Hamiltonian matrix is not explicitly constructed),
and an efficient parallel implementation of the matrix-vector operation is used based on the Sylvester 
algorithm.\cite{kendrick99,sylvesterHayes1994}
The dimension of 5D surface function Hamiltonian can be large especially for non-zero $J$.
However, the size of this matrix is significantly reduced by using the powerful
Sequential Diagonalization Truncation (SDT) technique.\cite{kendrick99,kendrickHD2GPJ5,bacicSDT1988,bacicSDT1990}
Furthermore, only the user specified $n$ lowest energy solutions are explicitly computed.
The surface functions are independent of the collision energy and only have to be computed once for a given PES.
However, they must be computed at each value of $\rho_\xi$ and for each value of total angular momentum $J$, inversion parity $\pm$, 
and exchange symmetry (for systems with identical nuclei).
All of these calculations are typically distributed over a large number of processors.

Once all of the surface functions have been computed and stored to disk, the appropriate potential coupling and overlap
matrices are computed and then used to solve the one-dimensional (1D) radial coupled-channel equation in $\rho$.
The coupled-channel equation in $\rho$ is solved using Johnson's log-derivative method.\cite{johnsonLOG,johnsonLOGNumerov}
The $n\times n$ log-derivative matrix is propagated from small $\rho$ to a user specified matching
distance $\rho_{\rm match}$ where the projection from APH to Delves coordinates is performed.
The log-derivative propagation is performed by sub-dividing each sector into several steps
with a uniform spacing in $\rho$ (typically $10$ - $50$ sub-steps within each sector are used depending 
upon the local de Broglie wavelength\cite{kendrickHD2GPJ5}).
Each propagation step requires several matrix-inversions which are performed using an efficient
LAPACK linear solver routine.\cite{LAPACK1999}
For large sets of coupled channels ($n>2000$) the parallel ScaLAPACK library can be used.\cite{SCALAPACK1997}
At the boundaries between the sectors, the log-derivative matrix is transformed to the new
basis centered at the next sector using the appropriate previously computed overlap matrix for the two sectors.
At the matching point $\rho_{\rm match}$ the log-derivative matrix is transformed to the
Delves basis using the overlap matrix between the APH surface functions and Delves channel functions computed at $\rho_{\rm match}$.
The Delves channel functions are computed using an efficient 1D Numerov\cite{johnsonLOGNumerov} propagator which computes a user specified
number of accurate vibrational solutions using Delves hyperspherical coordinates centered in each diatomic channel.\cite{parker2002}
The matching distance $\rho_{\rm match}$ is chosen to be large enough so that the diatomic channels are well separated 
and their solutions can be computed independently (i.e. the coupling and overlap between the different Delves channel basis functions can be ignored).
The vibrational manifold is computed up to a user specified energy for each diatomic rotational ($j$) and orbital angular momentum ($l$) quantum
number compatible with the specified total angular momentum ($J$), inversion parity ($\pm$) and particle exchange symmetry.
The log-derivative propagation is continued for $\rho_{\rm match}< \rho \le \rho_{\rm final}$ using the same techniques 
discussed above but now the potential coupling and overlap matrices have been computed using the Delves hyperspherical 
coordinates centered in each diatomic channel.
At the final asymptotic value of $\rho=\rho_{\rm final}$, the Delves log-derivative matrix is transformed to Jacobi coordinates
and analytic expressions for the asymptotic scattering solutions are used to compute the reactance ${\bf K}$  matrix and 
finally the full scattering ${\bf S}$ matrix.\cite{Pack87,kendrickHD2GPJ5}
The scatting matrix is computed at each specified energy and contains all asymptotically open initial and final 
channels labeled by the quantum numbers $v$ (vibrational), $j$ (rotational), and $l$ (orbital angular momentum)
for each diatomic channel ($\tau$).
From the scattering matrix the rotationally resolved (and also $m_j$ resolved), vibrationally resolved, and 
total cross sections and reaction rate coefficients can be computed as a function of collision energy 
(and scattering angle for the DCSs).

Extensive convergence studies were performed for each ultracold isotopic hydrogen exchange reaction.  
The primary convergence parameters associated with the 5D APH surface function solutions are 
$l_{\rm max}$, $m_{\rm max}$ and $^{\rm 1D}E_{\rm cut}$.
The positive integers $l_{\rm max}$ and $m_{\rm max}$ determine the number of basis functions 
in the hyperspherical angles $\theta$ and $\phi$, respectively.\cite{kendrick99,kendrickHD2GPJ5}
The $^{\rm 1D}E_{\rm cut}$ specifies the maximum energy of the $1D$ solutions kept during the SDT procedure.\cite{kendrick99,kendrickHD2GPJ5}
These parameters are optimized for different ranges in $\rho< \rho_{\rm match}$.
As $\rho$ increases, larger values are required for $l_{\rm max}$ and $m_{\rm max}$ due to the localization
of the surface functions in each diatomic channel.
The primary convergence parameters associated with the 1D Numerov
solution for the Delves channel vibrational functions for $\rho>\rho_{\rm match}$ and 
the asymptotic Jacobi channel vibrational solutions at $\rho=\rho_{\rm final}$ 
are the number of propagation steps and the initial and final values of the Delves or Jacobi coordinate.
The primary convergence parameters for the log-derivative propagation in the APH and Delves 
regions are the number of channels $n_{\rm APH}$ and $n_{\rm Delves}$, respectively.
Specific values for all of these parameters are given below.
The {\it ab initio} ground electronic state BKMP2\cite{BKMP2} PES was used in all of the calculations reported here.
Our previous ultracold quantum reactive scattering calculations using the Mielke\cite{MielkePES} PES gave similar results.\cite{Kendrik15_2}

For the HD$_2$ system the hyperradius $\rho$ was discretized within the APH region into $54$ logarithmically spaced 
sectors between $\rho=1.9\,{\rm a_o}$ and $\rho_{\rm match}=7.03\,{\rm a_o}$.
The centers of the sectors are given by $\rho_\xi= \rho_{\xi-1}(1 + \Delta\rho)$ where $\Delta\rho = 0.025\,{\rm a_o}$.
The optimal surface function basis sets were 
$l_{\rm max} =103,115,123,135,143$ and $m_{\rm max} = 190,214,232,250,274$ which correspond to the following ranges in $\rho$:
$1.9\le \rho \le 2.89$, $2.89 < \rho \le 3.61$, $3.61 < \rho \le 4.51$, $4.51 < \rho \le 5.63$, and $5.63 < \rho \le 7.03\,{\rm a_o}$, respectively.
The energy cut-off $^{1D}E_{\rm cut}$ used in the SDT procedure varied for each value of $\rho$. 
Representative values at the center of each of the five ranges in $\rho$ listed above are
$39.0$, $22.1$, $16.25$, $13.0$, and $10.4\,{\rm eV}$, respectively.
Before SDT truncation, the dimensions of the total $J=0$ surface function Hamiltonian matrix for each of the five basis sets are
$39\,624$, $49\,764$, $57\,660$, $68\,136$ and $79\,056$, respectively.
The maximum dimensions of the SDT truncated matrices for $J=0$ within each range in $\rho$ are
$10\,034$, $11\,706$, $13\,623$, $16\,646$, and $18\,828$, respectively.
For non-zero $J$ the dimensions of these matrices scale as $J+1$ (i.e., for $J=4^+$ the dimensions are $5$ times larger).
The solutions for even and odd exchange symmetry are projected out and propagated separately.
The number of channels used in the APH propagation region for each $J^p$ and a given exchange symmetry (even or odd) were
$300$, $600$, $900$, $1200$ and $1500$ for $J=0$, $1^-$, $2^+$, $3^-$ and $4^+$, respectively.
We note that for initial diatomic states in the ground $j=0$ rotational state, only the ${\bf S}$ matrices with even $J + p$ contribute to the cross sections.
At $\rho=\rho_{\rm match}=7.03\,{\rm a_o}$ the log-derivative matrices were transformed to the Delves channel functions using the overlap matrix between the APH and Delves surface functions.
The log-derivative propagation was continued using a uniform grid in $\rho$ between $7.03 < \rho \le 50.0\,{\rm a_o}$ with a sector spacing of $0.2\,{\rm a_o}$.
The number of channels propagated in the Delves region for each $J^p$ and a given exchange symmetry were
$300$, $590$, $874$, $1144$, and $1426$, respectively.
The Delves channel vibrational functions were computed using a 1D Numerov propagator with $n_{\rm steps}=6000$ points between $r=0.175$ and $6.0\,{\rm a_o}$ 
for each of the diatomic channels D$_2$ and HD.
These wave functions were down-sampled to $400$ points for use in the numerical quadratures for the various overlap and potential coupling calculations.
The asymptotic Jacobi 1D Numerov propagation used $n_{\rm steps}=5000$ between $r=0.1$ and $6.0\,{\rm a_o}$ and were down sampled to $500$ points.
The APH and Delves log-derivative propagation was carried out for $40$ logarithmically spaced collision energies between $1.16\,\mu{\rm K}$ ($10^{-10}\,{\rm eV}$) and
$100\,{\rm K}$ ($8.6\,{\rm meV}$) for vibrationally excited D + HD($v=4$, $j=0$) (i.e., total energy $\approx 1.91\,{\rm eV}$)
and H + D$_2$($v=4$, $j=0$) (i.e., total energy $\approx 1.59\,{\rm eV}$).
The various basis set parameters and number of coupled channels were optimized to give reliable scattering results over the entire collision energy range
for H/D collisions with the vibrationally excited HD($v=4$, $j=0$) and D$_2$($v=4$, $j=0$).

\section{Results and Discussion}
\label{sect3}
The quantum reactive scattering results for collisions of D with vibrationally 
excited HD($v=4$, $j=0$), and H with vibrationally excited D$_2$($v=4$, $j=0$)
for collision energies between $1\mu\,{\rm K}$ and $100\,{\rm K}$ will be presented in the following subsections \ref{sect3a} - \ref{sect3c}.
Rotationally resolved, vibrationally resolved, and total reaction rate coefficients will be presented for several products states as a function of collision energy.
The rate coefficients are computed from the integral cross sections ($\sigma_{if}$) using the expression $k_{if} = v\,\sigma_{if}$ where $v$ is the relative velocity
between the colliding atom and diatomic molecule. The labels $i$ and $f$ denote the initial $vjm_j$ and final $v'j'm'_{j'}$ states of the reactant and
product diatomic molecules, respectively.
In the present work, we average over the initial $m_j$ and sum over all final $m'_{j'}$ to obtain the $vj$ and $v'j'$ resolved results.
DCSs will also be presented as a function of both collision energy and scattering angle.
The results from two sets of calculations will be presented in each case: one set which includes the geometric phase (denoted by GP) and another set which does not
(denoted by No Geometric Phase (NGP)).
A new quantum interference mechanism is used to analyze and interpret the results.\cite{Kendrik15_1,Kendrik15_2,hazra2015JPC}

\subsection{D + HD $\to$ D + HD Reaction}
\label{sect3a}
\begin{figure}
\includegraphics[scale=0.6]{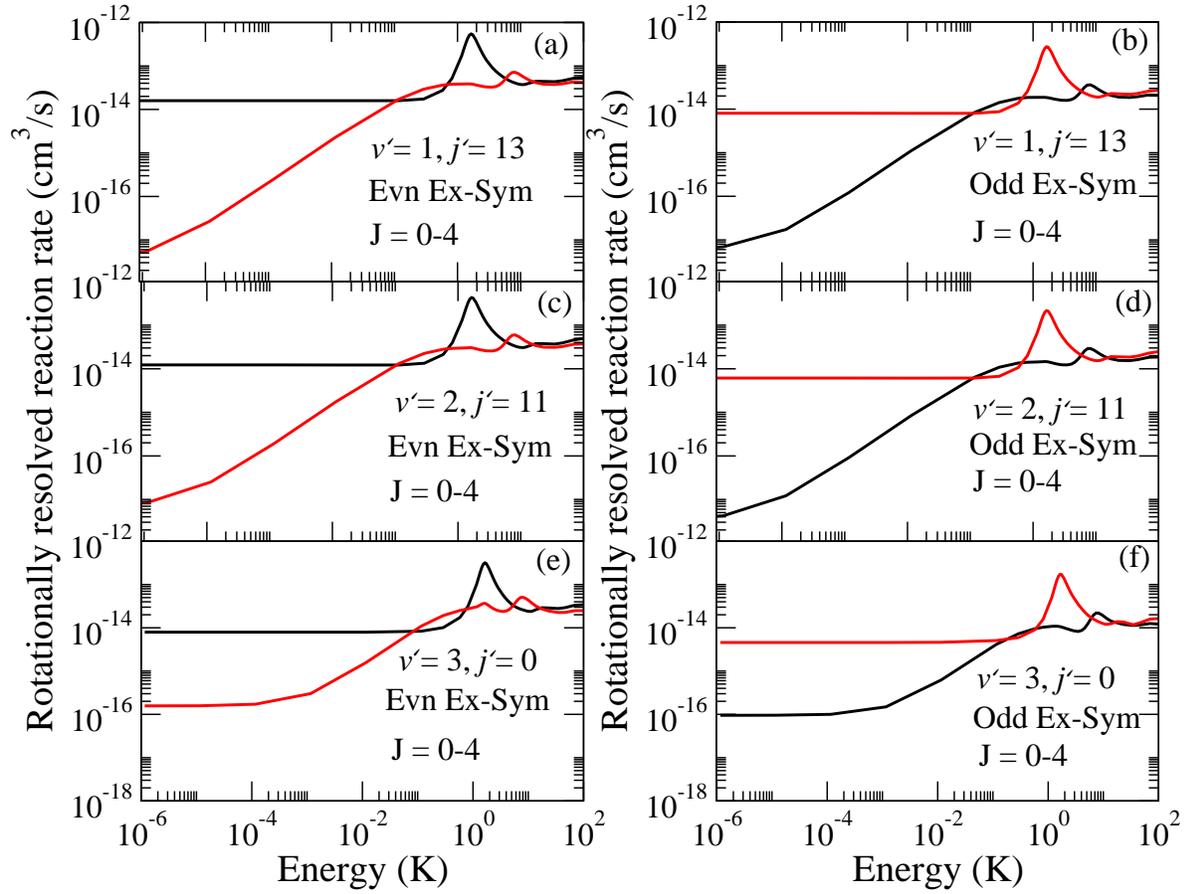}
\caption{
\begin{small} {Rotationally resolved rate coefficients for the D + HD($v=4$, $j=0$) $\to$ D + HD($v'$, $j'$) reaction are plotted
as a function of collision energy:  (a) and (b) $v'=1$, $j'=13$, (c) and (d) $v'=2$, $j'=11$, and (e) and (f) $v'=3$, $j'=0$.
The results for even exchange symmetry are plotted in the left panels (a), (c) and (e), and those for odd exchange symmetry are plotted in
the right panels (b), (d), and (f).  In all panels the red curves include the geometric phase (GP) and the black curves do not (NGP).  
The rates include all values of total angular momentum $J=0-4$.}
\end{small}
}
\label{fig2.1}
\end{figure}

Several representative rotationally resolved rate coefficients for the D + HD($v=4$, $j=0$) $\to$ D + HD($v'$, $j'$) reaction are plotted in Fig. \ref{fig2.1}
as a function of collision energy between $1\,\mu{\rm K}$ and $100\,{\rm K}$. The rate coefficients include all values of total angular momentum between $J=0-4$.
The GP was included using the vector potential approach for each value of $J$ and exchange symmetry.
The rate coefficients for even and odd exchange symmetry have been multiplied by the appropriate nuclear spin statistical factors of $2/3$ and $1/3$, respectively.
Since the identical D nuclei behave as spin 1 Bosons under permutation,\cite{HerzbergI} the total molecular wave function must always be symmetric with respect to this permutation.
Asymptotically the electronic wave function is symmetric with respect to a permutation of the identical D nuclei.\cite{HerzbergI,kendrickHD2GPJ}
Thus, the nuclear motion wave function of even (odd) exchange symmetry must be multiplied by the even (odd) nuclear spin function with the appropriate weight.
Fig. \ref{fig2.1} shows that for even (odd) exchange symmetry the GP suppresses (enhances) the ultracold rate coefficient.
For the HD($v'=1$, $j'=13$) and HD($v'=2$, $j'=11$) products the effect of the GP on the ultracold rate coefficients is over three orders of magnitude.
At higher energies shape resonances due to the $l=2$ and $3$ partial waves (see Fig. \ref{fig2.8}) are clearly visible near $1.8\,{\rm K}$ and $7\,{\rm K}$, respectively.
The results which include the GP predict that the prominent $l=2$ shape resonance occurs for odd exchange symmetry (the right panels in Fig. \ref{fig2.1}) 
but not for even exchange symmetry (the left panels in Fig. \ref{fig2.1}). 
In contrast, the less prominent $l=3$ shape resonance is predicted to occur for even exchange symmetry but not for odd exchange symmetry.
The calculations which ignore the GP give the opposite prediction for both resonances.
The predicted $l=2$ and $3$ shape resonance for odd and even exchange symmetry, respectively, provides an experimentally detectable signature 
of the GP effect in the D + HD($v=4$, $j=0$) reaction (assuming the appropriate nuclear spin state can be selected).
The huge suppression and enhancement of the ultracold reaction rate coefficient for even and odd exchange symmetry, respectively, provides
another experimentally detectable signal which might also be used to control the reaction (through selection of a particular nuclear spin state).
Tables \ref{table1} and \ref{table2} list several ultracold reaction rate coefficients for the D + HD($v=4$, $j=0$) $\to$ D + HD($v'$, $j'$) reaction for 
the even and odd exchange symmetries, respectively. Very large GP effects can be seen in many of the ultracold reaction rates coefficients for large $j'$. 
Most notable is the $4$ orders of magnitude suppression (enhancement) of the GP computed rate coefficient for $v'=0$ $j'=11$ even (odd) exchange symmetry.

The large GP effects in the ultracold vibrationally excited hydrogen exchange reactions are due to the efficient constructive or destructive 
interference which occurs between the non-reactive (no exchange) and reactive (exchange) processes (see Fig. 1 (a) in Ref. \cite{Kendrik15_2}).
The enhanced quantum interference is due to the unique properties associated with ultracold collisions:
(1) isotropic ($s$-wave) scattering, and (2) the relative phase between the two interfering scattering amplitudes often approaches an integral multiple of $\pi$.
These two properties enable maximum constructive or destructive interference to occur whenever the two interfering scattering amplitudes are similar in magnitude.
Specifically, let $f_1$ and $f_2$ denote the two interfering complex scattering amplitudes which we can write as $f_i = \vert f_i\vert\,\exp(i\,\delta_i)$.
The total scattering amplitude is $f_T = (f_1 + f_2)/\sqrt{2}$.  The cross sections are computed from the modulus of the total scattering amplitude
given by 
\begin{equation}
\vert\vert f_T \vert \vert = {1\over 2}\,( \vert f_1\vert^2 + \vert f_2\vert^2 + 2\,\vert f_1\vert\,\vert f_2\vert\,\cos\Delta)\, ,
\label{fmod}
\end{equation}
where the relative phase $\Delta=\delta_2 - \delta_1$. If the magnitudes of the scattering amplitudes are equal $\vert f_1\vert = \vert f_2\vert = f$,  then
Eq. \ref{fmod} reduces to $ \vert\vert f_T \vert \vert = f^2( 1 + \cos\Delta)$. Furthermore, if $\Delta = m\,\pi$ where $m$ is an integer, then we find
that $ \vert\vert f_T \vert \vert = 2\,f^2$ and $0$ for even and odd $m$, respectively.
In this case, maximum constructive (destructive) interference occurs for even (odd) $m$ which leads to an enhanced (suppressed) reaction rate.
As we will demonstrate below, for ultracold collisions the magnitudes of the two interfering scattering amplitudes are often similar and their relative phases often approach 
an integral multiple of $\pi$.
The effective quantization of the relative phase $\Delta$ between the two interfering scattering amplitudes can be understood in terms of a 1D potential well model for
which Levinson's theorem applies.\cite{Levinson1949,Jauch1957,Kazes1959,wright1965,newton1977,cheney1984,boya2007}
From Levinson's theorem we know that the scattering phase shift $\delta$ approaches an integral multiple of $\pi$ in the zero wave vector limit
(i.e.,  $\delta\to n\,\pi$ as $k\to 0$).
The integer $n$ corresponds to the number of bound states supported by the 1D potential well.
As the well depth is increased, the integer $n$ quickly jumps to $n+1$ as a continuum state drops into the well and becomes bound.
Applying this 1D model to our higher dimensional problem, 
the two interfering pathways sample a different region of the PES and therefore associated with each pathway is an effective 1D potential well with a different depth and/or width.
Thus, the number of bound states is different for each 1D potential well which gives rise to a phase which approaches a different integral multiple of 
$\pi$ (i.e. $\delta_i = n_i\,\pi$).
The difference between these two phases $\Delta = (n_2 - n_1)\,\pi = m\,\pi$ is still an integral multiple of $\pi$ which in general is either even or odd.
To our knowledge this is a new kind of quantum interference mechanism which is general and independent of the geometric phase.\cite{Kendrik15_1,Kendrik15_2,hazra2015JPC,hazra2016JPhysA}
If the two interfering pathways encircle a conical intersection, then the associated geometric phase leads to an additional $\pi$ phase shift which changes the sign on the
interference term in Eq. \ref{fmod}.
That is, for the conditions discussed above, Eq. \ref{fmod} becomes $\vert\vert f_T \vert \vert = f^2( 1 - \cos\Delta)$. 
In this case, maximum constructive (destructive) interference occurs for odd (even) $m$ which leads to an enhanced (suppressed) reaction rate.
Thus, we find that the opposite interference occurs when the geometric phase is included (i.e., the constructive interference becomes destructive and vice versa).
That is, the geometric phase controls the reaction.\cite{Kendrik15_1}
The complete suppression of the GP/NGP rate coefficients for even/odd exchange symmetries due to destructive interference makes the $s$-wave
contribution smaller than the $p$-wave contribution leading to notably different threshold behavior of the ultracold reaction rate coefficients in Fig. \ref{fig2.1} (a) - (f).

\begin{figure}
\includegraphics[scale=0.6]{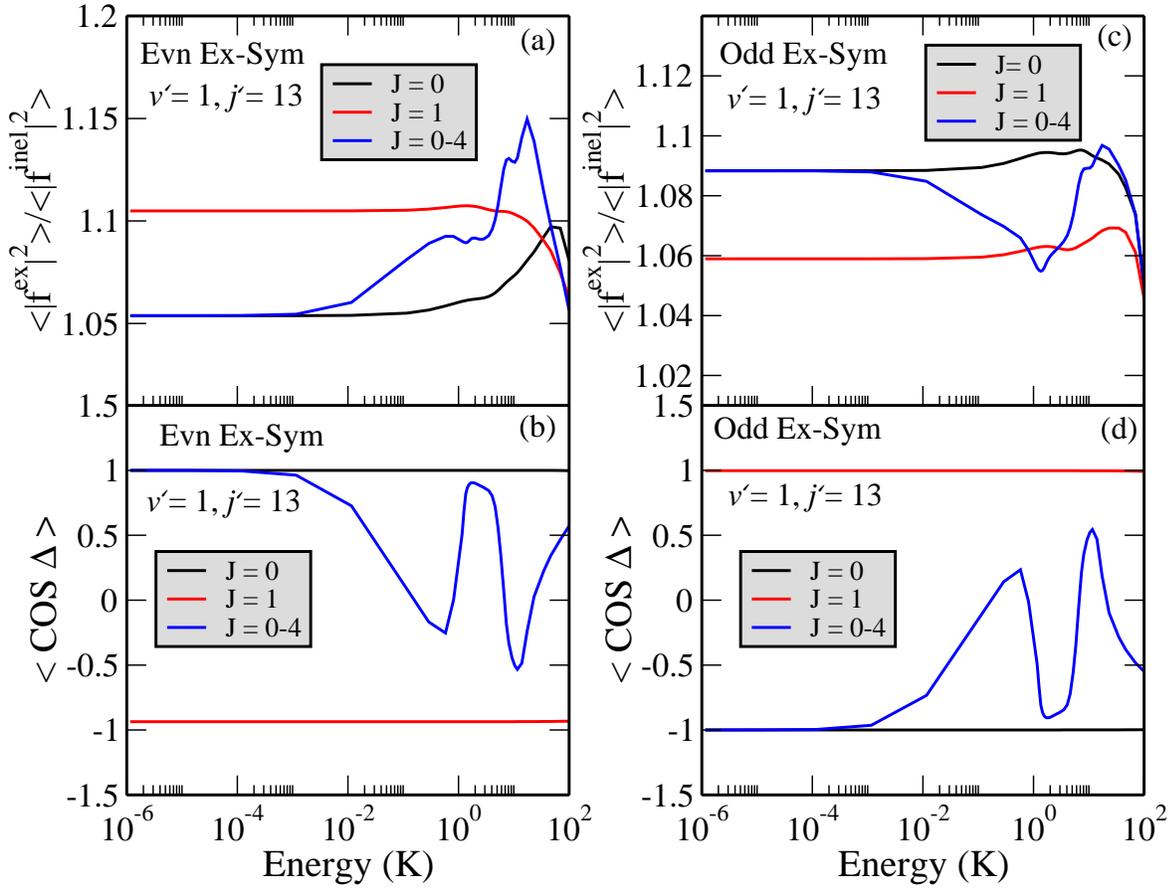}
\caption{
\begin{small} {Ratio of the average squared magnitude of the exchange and inelastic scattering amplitudes for the 
D + HD($v=4$, $j=0$) $\to$ D + HD($v'=1$, $j'=13$) reaction are plotted as a function of collision energy. 
The results in panels (a) and (c) correspond to even and odd exchange symmetry, respectively.
The average $\cos\Delta$ is plotted as a function of collision energy in panels (b) and (d) for even and odd exchange symmetry, respectively.
The results for $J=0$ are plotted in black, $J=1$ in red and summed over all $J=0-4$ in blue.}
\end{small}
}
\label{fig2.2}
\end{figure}

\begin{figure}
\includegraphics[scale=0.6]{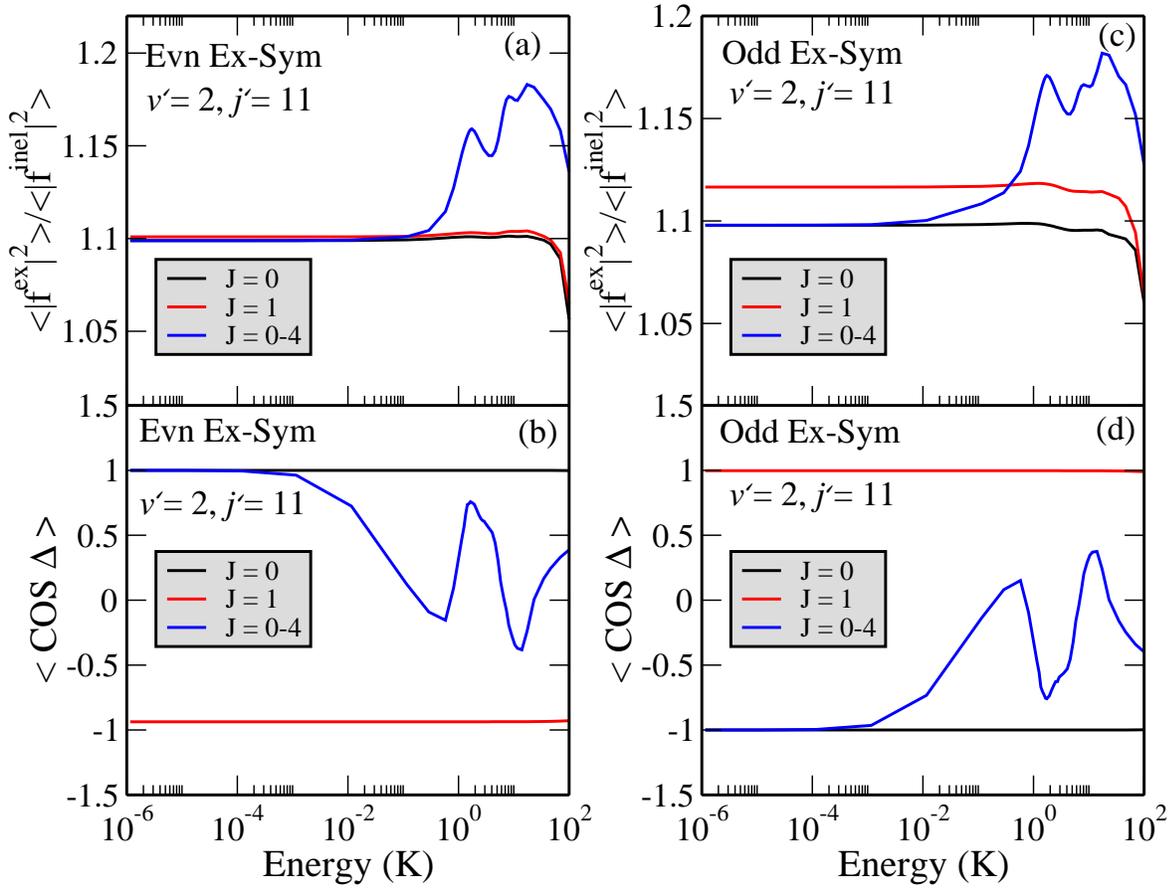}
\caption{
\begin{small} {Same as in Fig. \ref{fig2.2} but for $v'=2$, $j'=11$.}
\end{small}
}
\label{fig2.3}
\end{figure}

Figures \ref{fig2.2} - \ref{fig2.4} plot the ratio of the average squared magnitudes of the two interfering scattering amplitudes $f_1$ and $f_2$ 
denoted in these plots as $f^{\rm inel}$ (for inelastic/non-reactive) and $f^{\rm ex}$ (for exchange/reactive), respectively.
The two interfering scattering amplitudes are computed from the total scattering amplitudes $f^{NGP}$ and $f^{GP}$ via
$f^{\rm inel} = (f^{NGP} + f^{GP})/\sqrt{2}$ and $f^{\rm ex} = (f^{NGP} - f^{GP})/\sqrt{2}$.\cite{meadScatter,Althorpescience2005,Althorpe2006}
Also plotted in Figs. \ref{fig2.2} - \ref{fig2.4} is the average $\langle\cos\Delta\rangle$ for $m'_{j'} = j'$.
The averaging is with respect to the scattering angle $\theta$ and is defined as $\langle\,\rangle = (1/\pi)\int_0^\pi\,{\rm d}\theta$.
All of these quantities are plotted as a function of the collision energy between $1\,\mu{\rm K}$ and $100\,{\rm K}$.
The results plotted in Figs. \ref{fig2.2} - \ref{fig2.4} correspond to the $v'=1$, $j'=13$, 
$v'=2$, $j'=11$, and $v'=3$, $j'=0$ rates shown in Fig. \ref{fig2.1}, respectively.
In Figs. \ref{fig2.2} and \ref{fig2.3} panels (a) and (c) we see that the ratios are close to unity at ultracold collision energies.
For $J=0$ (black) the deviations from unity are approximately $0.06$, $0.09$, $0.1$ and $0.1$, respectively.
In Fig. \ref{fig2.4} panels (a) and (c) we see that the ratios for $J=0$ (black) deviate from unity at ultracold energies by approximately $0.7$ and $0.75$, respectively.
The larger differences between the magnitudes of the two scattering amplitudes reduce the interference effects.
This explains the smaller GP effect seen in the rate coefficient for $v'=3$, $j'=0$ plotted in Fig. \ref{fig2.1} panels (e) and (f).
The ratios for $J=1$ (red) and summed over all $J=0-4$ (blue) in Figs. \ref{fig2.2} and \ref{fig2.3} panels (a) and (c) 
are also close to unity with deviations between $0.05 - 0.2$ over the entire energy range.
The ratios for $J=1$ (red) and summed over all $J=0-4$ (blue) in Figs. \ref{fig2.4} are larger with deviations from unity up to $1.5$.
The $\langle\cos\Delta\rangle$ plotted in Figs. \ref{fig2.2} - \ref{fig2.4} panels (b) and (d) all approach $\pm 1$ at 
ultracold collision energies. 
This implies that the relative phase between the two interfering scattering amplitudes approaches an integral multiple of $\pi$.
Even and odd multiples of $\pi$ correspond to $\cos\Delta=1$ and $-1$, respectively.
Thus, for the NGP (GP) case $\cos\Delta=1$ and $-1$ leads to maximum constructive (destructive) or destructive (constructive) interference, respectively.
For even exchange symmetry in Figs. \ref{fig2.2} - \ref{fig2.4} panel (b), we see that $\langle\cos\Delta\rangle=1$ for $J=0$. 
Thus, constructive interference occurs for the NGP case and the ultracold NGP reaction rate coefficient is larger than the GP one in Fig. \ref{fig2.1} panels (a), (c), and (e).
For odd exchange symmetry in Figs. \ref{fig2.2} - \ref{fig2.4} panel (d), we see that $\langle\cos\Delta\rangle=-1$ for $J=0$. 
Thus, constructive interference occurs for the GP case and the ultracold GP reaction rate coefficient is larger than the NGP one in Fig. \ref{fig2.1} panels (b), (d), and (f).
In all cases, the $\langle\cos\Delta\rangle$ for $J=1$ have opposite sign than for $J=0$. 
This leads to opposite interference behavior in the $l=1$ partial wave contributions relative to $l=0$.
Thus, at higher collision energies where the $l=1$ partial wave begins to contribute, the differences between the NGP and GP rates decrease (see Fig. \ref{fig2.1}).
This trend continues for higher partial waves (with an occasional exception) and the interference behavior typically alternates with even and odd values of $l$ (recall $J=l$ here since $j=0$).
The $\langle\cos\Delta\rangle$ summed over $J=0-4$ deviate significantly from $\pm 1$ at higher collision energies and tend to oscillate about zero.
This explains the merging of the NGP and GP rates for collision energies above approximately $20\,{\rm K}$ where several partial waves contribute and effectively wash out the GP effect.
Some significant oscillations in $\langle\cos\Delta\rangle$ which approach $\pm 1$ are also seen at collision energies near $1.8\,{\rm K}$ and $7\,{\rm K}$ in Figs. 
\ref{fig2.2} - \ref{fig2.4} panels (b) and (d).
These energies correspond to the $l=2$ and $3$ shape resonances where significant interference occurs leading to large differences between the NGP and GP rates (see Fig. \ref{fig2.1}).

We have checked all open product states and without exception all of the ultracold reaction rates which exhibit a large GP effect 
have $\vert f^{ex}\vert^2/\vert f^{inel}\vert^2 \approx 1$ and $\cos\Delta \approx \pm 1$.
Fig. \ref{fig2.5} plots $\cos\Delta$ vs $\cos\Delta$ for all of the open product states $v'$, $j'$ for the 
D + HD($v=4$, $j=0$) $\to$ D + HD($v'$, $j'$) reaction at the ultracold collision energy of $1\,\mu{\rm K}$.
Both the even and odd exchange symmetries are plotted using black dots and red squares, respectively.
The majority of states are clustered near $\cos\Delta = \pm 1$ for which the GP effects are largest.
At higher collision energies, the distribution in $\cos\Delta$ spreads out along the diagonal (i.e., the relative phase between the two 
interfering scattering amplitudes is no longer ``quantized'').\cite{Kendrik15_2}
Gauge invariance was also verified by repeating the calculations with ${\rm m_A}=2$ in the expression for the vector potential ${\bf A}=-({\rm m_A}/2)\,\nabla\eta(x)$ where $\nabla$ is the gradient operator with respect to the nuclear coordinates $x$ and $\eta(x)$ is the azimuthal angle around the CI.\cite{meadtruhlar79,kendrick1996I,kendrick1996II,kendrickHD2GPJ5}
The GP (NGP) calculations correspond to odd (even) integers ${\rm m_A}$.  We used ${\rm m_A}=0$ and $1$ for the NGP and GP calculations, respectively.
The odd (even) values of ${\rm m_A}$ differ by a gauge transformation and give equivalent scattering results which include (do not include) the GP.
Thus, the results using ${\rm m_A}=2$ should be identical to those computed with ${\rm m_A}=0$ (NGP).
Fig. \ref{fig2.6} plots the same rates as in Fig. \ref{fig2.1} but also includes those computed with ${\rm m_A}=2$ (green squares).
To reduce the computational requirements, only the values of total angular momentum between $J=0-2$ are included in the gauge invariance check.
As expected, the reaction rate coefficients computed with ${\rm m_A}=0$ (the NGP curves plotted in black) are essentially identical to those computed with ${\rm m_A}=2$ (green squares).
Some differences are visible in panel (f) at ultracold energies but the magnitude of this rate coefficient is very small.
The gauge invariance check presented in Fig. \ref{fig2.6}  provides additional confirmation that the scattering results are well converged and that the large differences 
observed between the NGP (black) and GP (red) reaction rate coefficients are real and due entirely to the GP.

Fig. \ref{fig2.7} plots the vibrationally resolved rate coefficients for the D + HD($v=4$, $j=0$) $\to$ D + HD reaction summed over all final $j'$ states
as a function of collision energy between $1\,\mu{\rm K}$ and $100\,{\rm K}$. The rate coefficients include all values of total angular momentum between $J=0-4$.
In panels (a) and (c), the solid and dashed curves correspond to $v'=0$ and $1$, respectively.
In panels (b) and (d), the solid and dashed curves correspond to $v'=2$ and $3$, respectively.
The results for even exchange symmetry are presented in the left two panels (a) and (b), and the odd exchange symmetry results are presented in the right two panels (c) and (d).
In all cases, significant GP effects ($1$ to $3$ orders of magnitude) remain in the vibrationally resolved rates.
The $l=2$ and $3$ shape resonances are also clearly visible near $1.8\,{\rm K}$ and $7\,{\rm K}$, respectively.
Fig. \ref{fig2.8} plots the total rate coefficients for the D + HD($v=4$, $j=0$) $\to$ D + HD reaction summed over all final $v'$ $j'$ states
as a function of collision energy.
The results for even and odd exchange symmetry are plotted in panels (a) and (b), respectively.
The thick solid curves include all values of total angular momentum between $J=0-4$.
The individual contributions from each value of $J$ are plotted using thin curves:  solid $J=0$, dashed $J=1$, dot-dashed $J=2$, dotted $J=3$, and double-dot dashed $J=4$.
The $l=2$ and $3$ shape resonances are clearly visible near $1.8\,{\rm K}$ and $7\,{\rm K}$, respectively.
The effect of the GP on the total ultracold rate coefficients remains significant.
The total rate which includes the GP is suppressed (enhanced) by over an order of magnitude for even (odd) exchange symmetry.
In addition the $l=2$ ($l=3$) shape resonance is predicted to occur in the total rate for odd (even) exchange symmetry.
The opposite result is predicted if the GP is ignored.
Summing the total rate coefficients for the even and odd exchange symmetries in Fig. \ref{fig2.8} gives the total rate coefficient plotted in Fig. \ref{fig2.9} (as discussed
above for Fig. \ref{fig2.1} all of the rate coefficients include the appropriate nuclear spin statistical factors).
Due to the opposite behavior of the GP effects between the even and odd exchange symmetries (see Fig. \ref{fig2.8}), the GP effects largely cancel out in the
total rate when summed over both exchange symmetries.
Some small differences can be seen in the $l=2$ and $3$ shape resonances near $1.8\,{\rm K}$ and $7\,{\rm K}$, respectively.
The GP reduces the magnitude of the $l=2$ resonance and enhances the $l=3$ resonance.
The GP effect on the total ultracold reaction rate coefficient is now only a factor of $2$.
Thus, if a specific nuclear spin state can be selected, then the vibrationally resolved and total rate coefficients at ultracold collisions
also provide strong experimentally measurable signatures of the GP effect in the D + HD($v=4$, $j=0$) $\to$ D + HD reaction (see red curves in Figs. \ref{fig2.7} and \ref{fig2.8}).
If both nuclear spin states are present, then the GP effects tend to cancel out between the contributions from the even and odd exchange symmetries (see Fig. \ref{fig2.9})
which significantly reduces the possibility of experimental detection.

The DCS for the D + HD($v=4$, $j=0$) $\to$ D + HD($v'=3$, $j'=0$) reaction is plotted as function of collision energy and scattering angle in Fig. \ref{fig2.10}.
The DCS is plotted for collision energies between $1\,\mu{\rm K}$ and $20\,{\rm K}$ for which the cross section is well converged with respect to the partial wave sum.
Thus, the significant differences observed in the oscillatory structure between the NGP and GP DCSs (both in energy and scattering angle) are due entirely to the GP.
The isotropic scattering at ultracold collision energies due to the single $l=0$ partial wave ($s$-wave) is clearly visible.
The GP effect on the DCS is suppressed (enhanced) for even (odd) exchange symmetry.
The prominent $l=2$ shape resonance is clearly visible near $1.8\,{\rm K}$ and exhibits the expected oscillatory structure in the scattering angle (i.e., the three ``humps'').
The results which include the GP predict that this resonance occurs for odd exchange symmetry (panel b) while the NGP results give the opposite prediction (see panel (a)).
The less prominent $l=3$ shape resonance occurs near $7\,{\rm K}$.
In contrast to the $l=2$ shape resonance, the results which include the GP predict that the $l=3$ resonance occurs for even exchange symmetry (panel a) while the NGP 
results give the opposite prediction (see panel (b)).
Fig. \ref{fig2.11} plots the DCS for the D + HD($v=4$, $j=0$) $\to$ D + HD($v'=3$, $j'=0$) reaction as a function of scattering angle for two fixed collision energies
near the shape resonances seen in Fig. \ref{fig2.10} (only the results for odd exchange symmetry are plotted): $E_c = 1.6\,{\rm K}$ panel (a) and $E_c = 10.4\,{\rm K}$ panel (b).
At both energies, significant differences ($1$ to $3$ orders of magnitude) occur in the oscillatory structure between the DCSs computed with (red) and without (black) the GP.
At $1.6\,{\rm K}$ (panel (a)) the contribution from the $l=2$ shape resonance dominates the GP DCS and its associated angular dependence (i.e., $\vert\cos 2\theta\vert^2$)
is clearly visible.
In contrast, the NGP DCS exhibits the symmetric forward/backward $l=1$ angular behavior (i.e., $\vert\cos\theta\vert^2$) and its overall magnitude is suppressed.
At $10.4\,{\rm K}$ (panel (b)) the $l=3$ resonance dominates the NGP DCS which enhances its overall magnitude so that it is now comparable in magnitude to the GP DCS.
The $l=3$ angular dependence of the NGP DCS exhibits the expected $\vert\cos 3\theta\vert^2$ with three nodes at $30^\circ$, $90^\circ$, and $150^\circ$.
However, the $l=3$ resonance is suppressed in the GP DCS at $10.4\,{\rm K}$ (for odd exchange symmetry) so that it still exhibits the $l=2$ angular dependence seen in panel (a).
The results for even exchange symmetry (not plotted) are essentially the same as those plotted in Fig. \ref{fig2.11} except that the behavior of the GP (red) and NGP (black) curves
are reversed.
The large differences (nearly two orders of magnitude) between the DCS computed with and without the GP at ultracold collision energies 
provides a strong experimentally measurable signal for detecting GP effects in the D + HD($v=4$, $j=0$) $\to$ D + HD($v'=3$, $j'=0$) reaction.
At higher collision energies near $E_c = 1.6\,{\rm K}$ and $E_c = 10.4\,{\rm K}$ the $l=2$ and $3$ shape resonances provide additional 
experimentally detectable signatures of the GP effect.
The predicted oscillatory structure of the DCS in both energy and scattering angle is strongly dependent upon the GP and exchange symmetry
varying by up to three orders of magnitude.

Argand plots are presented in Fig. \ref{fig2.12} for the D + HD($v=4$, $j=0$) $\to$ D + HD($v'=3$, $j'=0$) reaction
for $J=l=0$ (panels (a) and (c)), and $J=l=2$ (panels (b) and (d)).
The NGP results for even exchange symmetry are plotted in the left panels (a) and (b), and the
GP results for odd exchange symmetry are plotted in the right panels (c) and (d).
The results for $J=0$ show a smooth non-resonant clockwise trajectory as the collision energy increases from 
$1\,\mu{\rm K}$ (near the origin of the plots) to $100\,{\rm K}$.\cite{kendrickDH2GPJ}
The energy range near the $l=2$ shape resonance occurring at $Ec=1.6\,{\rm K}$ is indicated by the red squares.
The results for $J=2$ show a counter-clockwise trajectory as the collision energy increases due to the $l=2$ shape resonance.\cite{kendrickDH2GPJ}
At higher collision energies, the larger non-resonant background contributions dominate and cause the trajectory to reverse direction
and move clockwise.
These loops or kinks in the argand trajectories are well known signatures of quantum resonances,\cite{arno,kendrickDH2GPJ}
and provide additional confirmation that the observed bumps in the rate coefficients and DCSs plotted in Figs. \ref{fig2.1}, \ref{fig2.6} - \ref{fig2.11}
are in fact due to quantum resonances.
A Lorentzian fit including background contributions was performed for the $l=2$ and $3$ shape resonances to more accurately determine their properties.
The fit was performed for the total reaction rate coefficients computed with the GP plotted in Fig. \ref{fig2.8}.
The $l=2$ shape resonance is predicted to occur for odd exchange symmetry (thick solid red curve in panel (b) of Fig. \ref{fig2.8})
and the $l=3$ shape resonance is predicted to occur for even exchange symmetry (thick solid red curve in panel (a) of Fig. \ref{fig2.8})
The resulting fits give a resonance energy for the $l=2$ ($l=3$) shape resonance of $E_c=1.68\,{\rm K}$ ($E_c=7.60\,{\rm K}$),
a width of $\Delta E = 0.854\,{\rm K}$ ($\Delta E = 5.71\,{\rm K}$), and 
the corresponding lifetime is $\tau = 35.8\,{\rm ps}$ ($\tau = 5.35\,{\rm ps}$).

\subsection{D + HD $\to$ H + D$_2$ Reaction}
\label{sect3b}
For most product states the GP effects in the D + HD($v=4$, $j=0$) $\to$ H + D$_2$($v'$, $j'$) reaction are significantly
smaller than those reported above in Sect. \ref{sect3a} for the D + HD($v=4$, $j=0$) $\to$ D + HD($v'$, $j'$) reaction.
This is primarily due the large difference in magnitude between the two interfering scattering amplitudes (the direct and looping pathways in this case, 
see Fig. 1 (b) in Ref.  \cite{Kendrik15_2}).  
The direct pathway typically dominates in the D + HD $\to$ H + D$_2$ reaction so that little interference occurs with the looping pathway.
There are a few notable exceptions which occur for large $j'$ and the rotationally resolved reaction rates coefficients for these are plotted in Fig. \ref{fig3.1}
as a function of collision energy between $1\,\mu{\rm K}$ and $100\,{\rm K}$. The rate coefficients include all values of total angular momentum between $J=0-4$.
The overall magnitudes of the reaction rate coefficients in Fig. \ref{fig3.1} are significantly smaller (by two orders of magnitude) 
relative to those plotted in Fig. \ref{fig2.1} for the D + HD($v=4$, $j=0$) $\to$ D + HD($v'$, $j'$) reaction.
However, the overall behavior is similar.
In particular, the GP rate is suppressed (enhanced) for even (odd) exchange symmetry and the prominent $l=2$ and $3$ shape resonances are clearly visible
near $2\,{\rm K}$ and $8\,{\rm K}$, respectively.
As in Fig. \ref{fig2.1}, the GP results in Fig. \ref{fig3.1} predict that the $l=2$ ($l=3$) shape resonance occurs for odd (even) exchange symmetry.
A Lorentzian fit including background contributions was performed for the GP predicted $l=2$ and $3$ shape resonances in Fig. \ref{fig3.1} panels (d) and (a), respectively.
The resonance energy, width and lifetime for the GP $l=2$ shape resonance in panel (d) are  $E_{\rm res} = 1.70\,{\rm K}$, 
$\Gamma = 0.865\,{\rm K}$ and $\tau = 35.3\,{\rm ps}$, respectively.
The resonance energy, width and lifetime for the GP $l=3$ shape resonance in panel (a) are  $E_{\rm res} = 7.68\,{\rm K}$, 
$\Gamma = 6.41\,{\rm K}$ and $\tau = 4.76\,{\rm ps}$, respectively.
The GP effects in the ultracold vibrationally resolved and total reaction rate coefficients (not plotted) are relatively small ($1.2$ - $1.5\times$) and ($1.3\times$), respectively.

Gauge invariance was verified for the results plotted in Fig. \ref{fig3.1} by performing a third calculation which included the
vector potential but with ${\rm m_A}=2$.  
The results for ${\rm m_A}_2=2$ are plotted in Fig. \ref{fig3.3} (green squares) for the same rates plotted in Fig. \ref{fig3.1}.
To reduce the computational expense, the gauge invariance check was restricted to total angular momentum between $J=0-2$.
The NGP (black) and ${\rm m_A}_2=2$ (green squares) are nearly identical for all collision energies. This confirms that the
results are well converged and that the differences between the GP (red) and NGP (red) results are real and due entirely to the GP.

The $J=l$ resolved (recall $J=l$ here since $j=0$) 
rate coefficients for the D + HD($v=4$, $j=0$) $\to$ H + D$_2$($v'=3$, $j'=10,11$) reactions are plotted in Fig. \ref{fig3.4}
as a function of collision energy between $1\,\mu{\rm K}$ and $100\,{\rm K}$. 
The two product states $v'=3$ $j'=10$ (panel a) and $v'=3$ $j'=11$ (panel b) correspond to the rates plotted in Fig. \ref{fig3.3}
panels (c) and (d), respectively.
As observed for the D + HD($v=4$, $j=0$) $\to$ D + HD($v'$, $j'$) reaction (see Fig. \ref{fig2.8}), the
ultracold $J=0$ NGP rate is larger (smaller) than the GP one for even (odd) exchange symmetry.
Also, the dominant contribution (i.e. whether the NGP or GP rate is largest) typically (but not always) 
alternates between even and odd values of $J$.\cite{kendrickHD2GPJ5,kendrickHD2GPJ,kendrickDH2GPJ,kendrickFeature2003}
The $l=2$ and $3$ shape resonances are clearly visible in both the NGP and GP $J$-resolved rate coefficients.
Due to constructive interference, the NGP $l=2$ shape resonance (near $2\,{\rm K}$) dominates for even exchange symmetry (panel a) while the GP $l=2$ shape resonance 
is masked by the larger GP $l=1$ background.
For odd exchange symmetry, the opposite behavior occurs and it is the GP $l=2$ shape resonance which dominates (the NGP $l=2$ shape resonance
is masked by the larger NGP $l=1$ background).
Similar but opposite symmetry behavior is observed for the weaker $l=3$ shape resonance near $8\,{\rm K}$.

\subsection{H + D$_2$ $\to$ D + HD Reaction}
\label{sect3c}
Similar to the D + HD($v=4$, $j=0$) $\to$ H + D$_2$($v'$, $j'$) reaction discussed above in Sect. \ref{sect3b},
the GP effects for the H + D$_2$($v=4$, $j=0$) $\to$ D + HD($v'$, $j'$) reaction are also 
much smaller than those reported for D + HD($v=4$, $j=0$) $\to$ D + HD($v'$, $j'$) in Sect. \ref{sect3a}.
Again, this is primarily due to the large difference in magnitude between the direct and looping scattering amplitudes.\cite{Kendrik15_2}
The direct pathway also dominates for the H + D$_2$ $\to$ D + HD reaction so that little interference occurs with the looping pathway.
There are only a few exceptions which occur for small $j'$ and the rotationally resolved reaction rate coefficients for these are plotted in Fig. \ref{fig4.1}.
The product state with the largest ultracold rate coefficient ($v'=0$, $j'=1$ in panel (a)) shows that the NGP rate is only slightly enhanced (by a factor of approximately $1.6$) relative to the GP rate. 
In contrast, it is the ultracold GP rate coefficient which is largest for the $v'=2$, $j'=5$ product state plotted in panel (b) (the ultracold GP rate is approximately $3.8$ times larger than the NGP rate).
The vibrationally resolved rate coefficient for $v'=0$ summed over all $j'$ is plotted in panel (c) for which the NGP ultracold rate coefficient is approximately $1.3$ times larger than the GP rate.
Very small GP effects ($<5\%$) were found in the other vibrationally resolved ultracold rate coefficients (not plotted).
No GP effect is observed in the total rate coefficient summed over all product $v'$ $j'$ in panel (d).
An $l=1$ partial wave shape resonance occurs in both the GP and NGP rate coefficients near $1\,{\rm K}$.
A Lorentzian fit including background contributions to the total rate plotted in panel (d) gives
a resonance energy, width and lifetime of $E_{\rm res} = 0.619\,{\rm K}$, $\Gamma = 0.824\,{\rm K}$ and $\tau = 37.1\,{\rm ps}$, respectively.

Gauge invariance was verified for the results plotted in Fig. \ref{fig4.1} by performing a third calculation which included the
vector potential but with ${\rm m_A}=2$.  
The results for ${\rm m_A}=2$ are plotted in Fig. \ref{fig4.2} (green squares) for the same rates plotted in Fig. \ref{fig4.1}.
To reduce the computational expense, the gauge invariance check was restricted to total angular momentum between $J=0-2$.
In panel (a), the NGP (black) and ${\rm m_A}=2$ (green squares) are essentially identical at ultracold collision energies but
show some small differences near the $l=1$ shape resonance.
Thus, we conclude that the slight suppression in the GP rate at ultracold energies is real but that the small differences between
the NGP and GP results near the $l=1$ shape resonance are not significant.
In panel (b) the NGP (black) and ${\rm m_A}=2$ (green squares) are in good overall agreement near the shape resonance but
show larger differences at ultracold energies.
In panels (c) and (d) good agreement is observed between the NGP (black) and ${\rm m_A}=2$ (green squares) at all collision energies.
The small differences between the NGP (black) and ${\rm m_A}=2$ (green squares) in Fig. \ref{fig4.2} panels (a) - (d)
indicate the level of convergence and also give an upper estimate for the uncertainty in the differences between the 
NGP (black) and GP (red) rate coefficients plotted in Fig. \ref{fig4.1}.
The GP results are computed with ${\rm m_A}=1$ and smaller ${\rm m_A}$ are typically better converged than larger ${\rm m_A}$.

\vfill\eject
\section{Conclusions}
\label{sect4}
The results of quantum reactive scattering calculations for the vibrationally excited 
D + HD($v=4$, $j=0$) $\to$ D + HD($v'$, $j'$), and D + HD($v=4$, $j=0$) $\leftrightarrow$ H + D$_2$($v'$, $j'$)
reactions were presented for collision energies between $1\,\mu{\rm K}$ and $100\,{\rm K}$.
For vibrationally excited reactants ($v>3$), these reactions become barrierless and proceed over
an effective potential well (along the vibrational adiabat).\cite{truhlar1970,miller1980,truhlar1996,jankunas_PNAS2014}
Thus, significant reactivity occurs even for ultracold collision energies.\cite{Simbotin2011,Simbotin2014,Simbotin2015}
An accurate full dimensional time-independent coupled-channel approach based on hyperspherical coordinates was used.\cite{Pack87,kendrick99,parker2002}
The calculations were performed using the {\it ab initio} BKMP2 PES for the ground electronic state of H$_3$.\cite{BKMP2}
All values of total angular momentum between $J=0-4$ are included.
The general vector potential approach was used to include the GP and
two sets of scattering results were presented for each reaction: one which included the GP
and one which did not.\cite{meadtruhlar79,kendrick1996I,kendrick1996II,kendrickHD2GPJ5}
Very large GP effects (up to $3$ orders of magnitude) were reported in the ultracold reaction rate coefficients.
The GP effects are typically largest for rotationally resolved rate coefficients but large
effects remain in the vibrationally resolved and total rates in many cases.
Significant GP effects were also reported in the DCSs which alter both the magnitude and oscillatory 
structure of the DCSs.
Several partial wave shape resonances occur for higher collision energies
between $0.5$ and $10\,{\rm K}$.
The GP is shown to dramatically alter the predicted resonance spectrum.
The large GP effects are explained in terms of a new quantum interference mechanism
which originates from the unique properties associated with ultracold collisions.\cite{Kendrik15_1,Kendrik15_2,hazra2015JPC,hazra2016JPhysA}
This novel quantum interference mechanism can lead to maximum constructive or destructive interference 
for ultracold collisions which effectively turns the reaction on or off, respectively.
The GP controls whether the interference is constructive or destructive (i.e., it acts like a quantum switch).

Rotationally resolved rate coefficients for the 
D + HD($v=4$, $j=0$) $\to$ D + HD($v'$, $j'$) reaction show
very large GP effects (up to $3$ orders of magnitude) at ultracold collision energies.
The GP computed rates are smaller (larger) than the NGP rates for even (odd) symmetry.
At higher collision energies, the GP computed rate coefficients predict
the appearance of a prominent $l=2$ shape resonance at $1.68\,{\rm K}$
for odd exchange symmetry and a weaker $l=3$ shape resonance is predicted to occur
at $7.60\,{\rm K}$ for even exchange symmetry.
The NGP computed rate coefficients predict the opposite symmetry
behavior (i.e., the $l=2$ ($l=3$) shape resonance occurs for even (odd) exchange symmetry).
Large GP effects (up to $3$ orders of magnitude) also remain in the vibrationally resolved 
ultracold rate coefficients, but are reduced ($25$ to $80\times$) in the total 
rate coefficients for each exchange symmetry.
The predicted $l=2$ and $3$ shape resonances are also present in the vibrationally resolved 
and total rate coefficients for each exchange symmetry.
Significant cancellation of the GP effects occurs when the rate coefficients for
each exchange symmetry are added to obtain a total rate.

The GP effects in the ultracold rate coefficients for the 
D + HD($v=4$, $j=0$) $\to$ H + D$_2$($v'$, $j'$)
and H + D$_2$($v=4$, $j=0$) $\to$ D + HD($v'$,$j'$) reactions
are typically much smaller than those reported for 
the D + HD($v=4$, $j=0$) $\to$ D + HD($v'$,$j'$) reactions.
This is primarily due to the differences in the two interfering pathways.
For D + HD $\leftrightarrow $ H + D$_2$ the two interfering reaction
pathways are the direct and looping pathways.
The direct pathway typically dominates the scattering process so that little
interference with the looping pathway occurs and therefore little or no GP effects appear.
In contrast, it is the non-reactive (inelastic) and reactive (exchange) pathways
that interfere in the D + HD $\to$ D + HD reactions.
Both of these pathways contribute comparably to the scattering process so that
large interference effects can occur and therefore large GP effects can appear.
Nevertheless, some significant GP effects (up to $2$ orders of magnitude)
are found to occur in the ultracold D + HD($v=4$, $j=0$) $\to$ H + D$_2$($v'$, $j'$)
reaction rate coefficients for a few values of large $j'$ (where both the direct and looping scattering 
amplitudes have similar magnitudes).

We have demonstrated that the unique properties associated with ultracold collisions can lead to significantly
enhanced quantum interference effects.  For barrierless reactions which proceed over
a potential well, the interference between two contributing reaction pathways can 
approach the maximum allowed values for constructive or destructive interference
(i.e., effectively turning the reaction on or off).
This new quantum interference mechanism is general and is expected to occur in a large
number of ultracold chemical reactions.\cite{Kendrik15_1,Kendrik15_2,hazra2015JPC,hazra2016JPhysA}
If the system exhibits a conical intersection, then the associated GP alters the relative 
sign of the interference term between the two interfering pathways which encircle 
the conical intersection.  Thus, the GP controls the interference and therefore
the outcome of the ultracold chemical reaction. 
The large GP effects reported here for the vibrationally excited hydrogen exchange
reactions provide new and exciting opportunities for the experimental confirmation
of the GP effect in a chemical reaction.
We hope that these results stimulate additional theoretical and experimental 
investigation of the hydrogen and other chemical reactions in the largely unexplored
ultracold/cold energy regime.


\subsection*{Acknowledgments}
\begin{small}
BKK acknowledges that part of this work was done under the auspices of
the US Department of Energy under Project No. 20140309ER of the Laboratory Directed
Research and Development Program at Los Alamos National Laboratory. Los Alamos National 
Laboratory is operated by Los Alamos National Security, LLC, for the National Security
Administration of the US Department of Energy under contract DE-AC52-06NA25396.
The UNLV team acknowledges support from the Army Research Office, MURI
grant No.~W911NF-12-1-0476 and the National Science Foundation, grant No.~PHY-1505557.
\end{small}

\vfill\eject


\vfill\eject



\begin{figure}
\includegraphics[scale=0.6]{Fig2_4.eps}
\caption{
\begin{small} {Same as in Fig. \ref{fig2.2} but for $v'=3$, $j'=0$.}
\end{small}
}
\label{fig2.4}
\end{figure}

\begin{figure}
\includegraphics[scale=0.6]{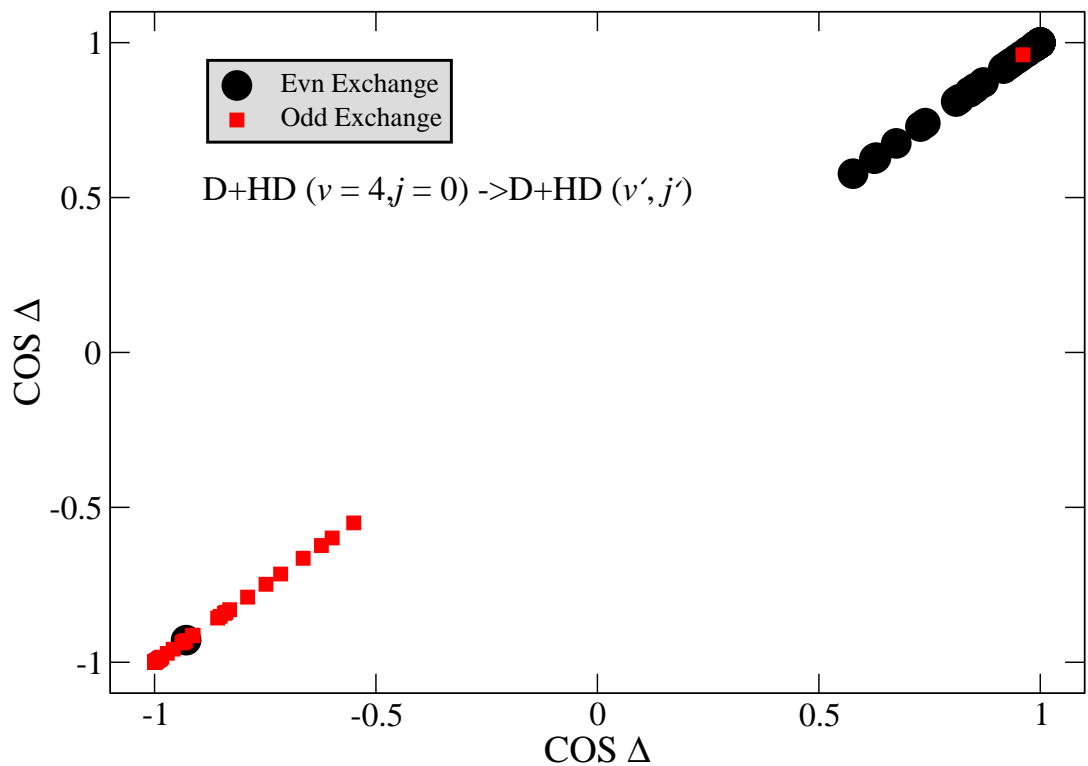}
\caption{
\begin{small} {The $\cos\Delta$ vs $\cos\Delta$ plot for each product $v'$ $j'$ state for the 
D + HD($v=4$, $j=0$) $\to$ D + HD($v'$, $j'$) reaction at the ultracold collision energy of $1\,\mu{\rm K}$.
The results for even and odd exchange symmetry are plotted with black dots and red squares, respectively.}
\end{small}
}
\label{fig2.5}
\end{figure}

\begin{figure}
\includegraphics[scale=0.6]{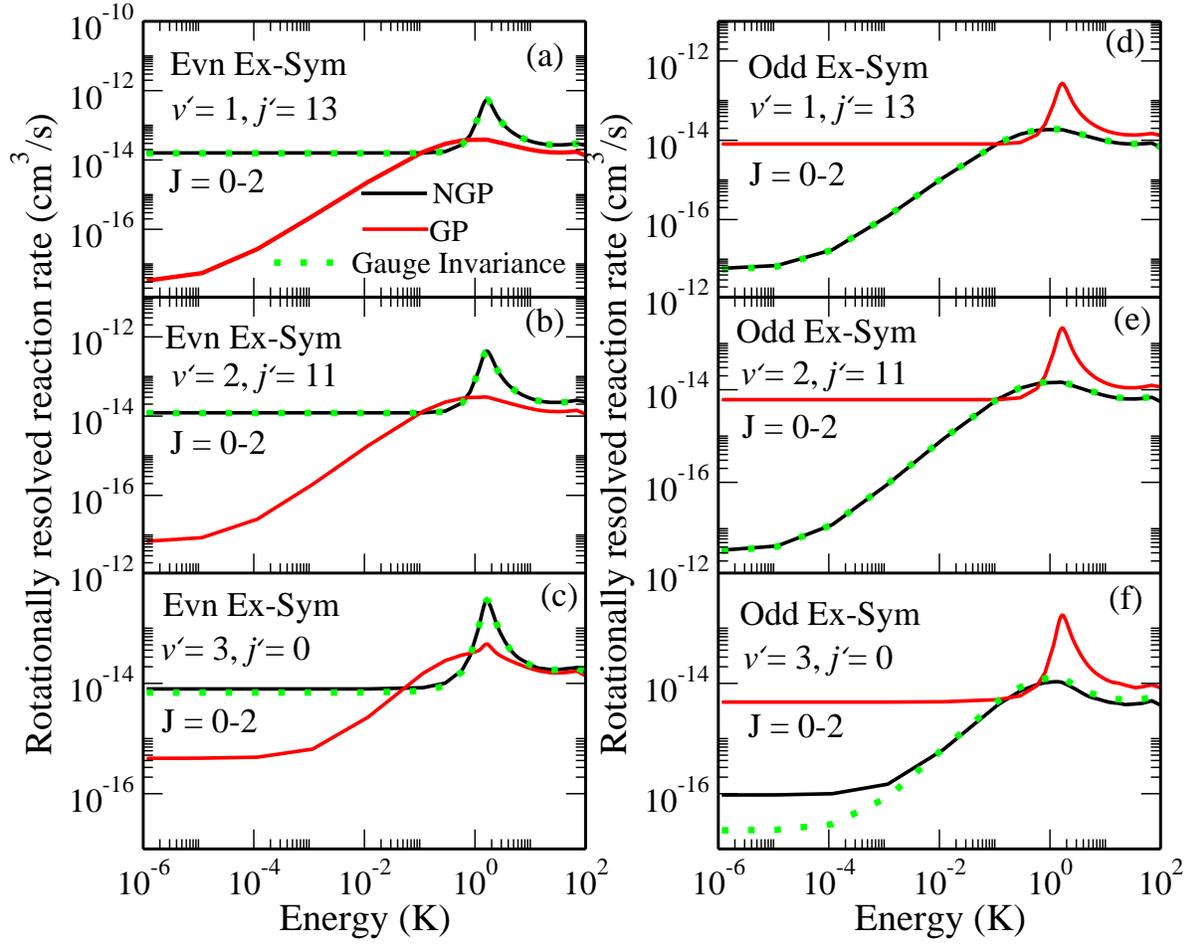}
\caption{
\begin{small} {Gauge invariance check (green squares) for the same product states plotted in Fig. \ref{fig2.1}. Here
only the total values of angular momentum between $J=0-2$ are included.}
\end{small}
}
\label{fig2.6}
\end{figure}

\begin{figure}
\includegraphics[scale=0.6]{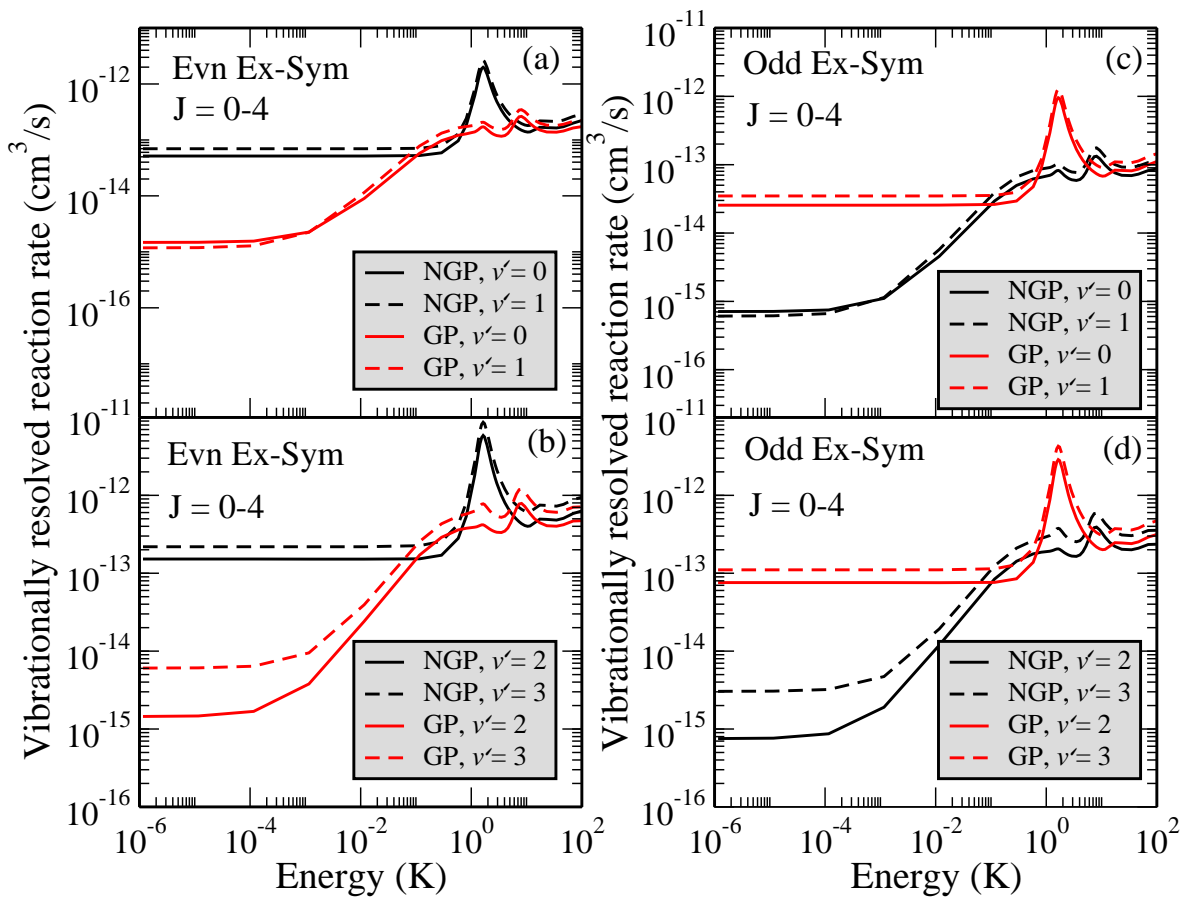}
\caption{
\begin{small} {Vibrationally resolved rate coefficients for the D + HD($v=4$, $j=0$) $\to$ D + HD($v'$) reaction are plotted
as a function of collision energy:  (a) and (c) $v'=0$ solid, $v'=1$ dashed; (b) and (d) $v'=2$ solid, $v'=3$ dashed.
The left two panels (a) and (b) are even exchange symmetry and the right two panels (c) and (d) are odd exchange symmetry.
In all panels the black and red curves correspond to NGP and GP, respectively. 
The rates include all values of total angular momentum $J=0-4$.}
\end{small}
}
\label{fig2.7}
\end{figure}

\begin{figure}
\includegraphics[scale=0.6]{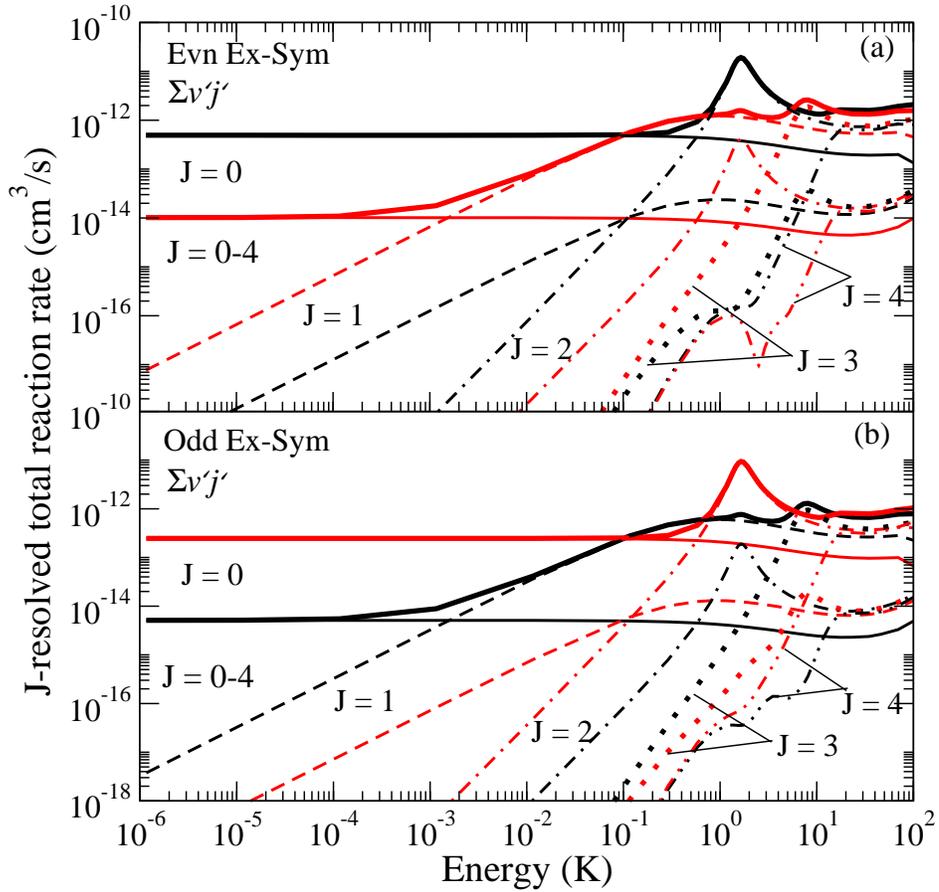}
\caption{
\begin{small} {The total rate coefficients for the D + HD($v=4$, $j=0$) $\to$ D + HD reaction are plotted
as a function of collision energy:  (a) even exchange symmetry and (b) odd exchange symmetry.
The thick solid curves include all values of total angular momentum $J=0-4$.
Thin solid curves $J=0$, dashed curves $J=1$, dot dashed $J=2$, dotted $J=3$, and double-dot dashed $J=4$.
In all cases the black and red curves correspond to NGP and GP, respectively.}
\end{small}
}
\label{fig2.8}
\end{figure}

\begin{figure}
\includegraphics[scale=0.6]{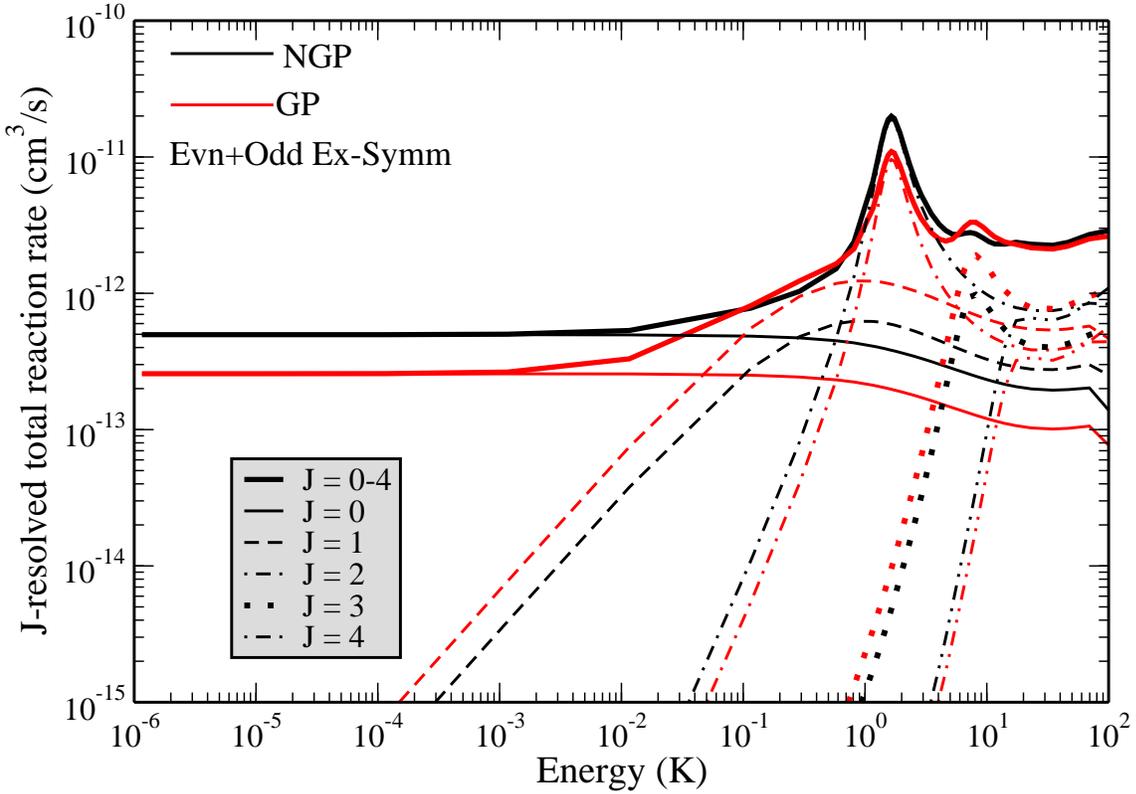}
\caption{
\begin{small} {The total rate coefficients for the D + HD($v=4$, $j=0$) $\to$ D + HD reaction are plotted
as a function of collision energy including contributions from both even and odd exchange symmetry.
The thick solid curves include all values of total angular momentum $J=0-4$.
Thin solid curves $J=0$, dashed curves $J=1$, dot dashed $J=2$, dotted $J=3$, and double-dot dashed $J=4$.
In all cases the black and red curves correspond to NGP and GP, respectively.}
\end{small}
}
\label{fig2.9}
\end{figure}

\begin{figure}
\includegraphics[scale=0.35]{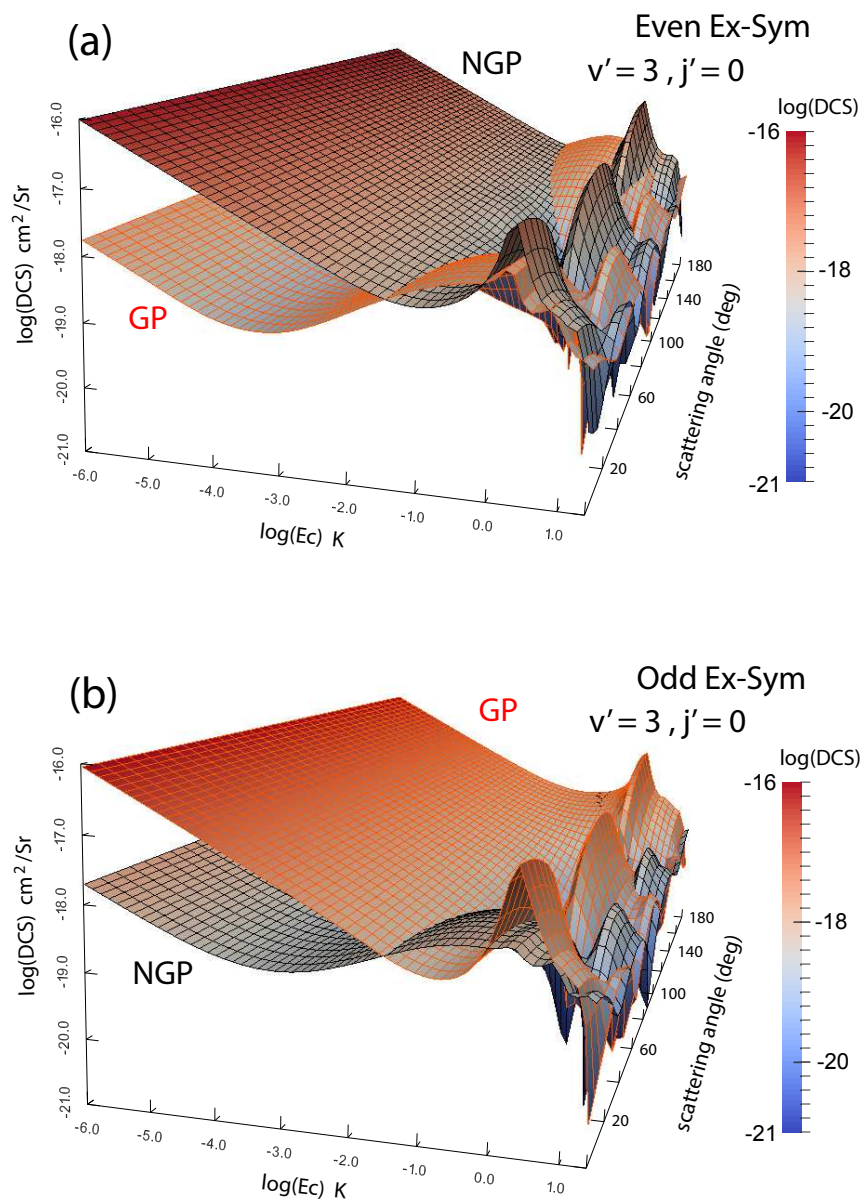}
\centering
\begin{tabular}{cc}
\end{tabular}
\caption{
\begin{small} {The DCS is plotted as a function of collision energy and scattering angle
for the D + HD($v=4$, $j=0$) $\to$ D + HD($v'=3$, $j'=0$) reaction.
Panel (a) is even exchange symmetry and (b) is odd exchange symmetry.
The DCS plotted with the red mesh includes the geometric phase (GP) while the one with the black mesh does not (NGP).
The results include all values of total angular momentum $J=0-4$.}
\end{small}
}
\label{fig2.10}
\end{figure}

\begin{figure}
\includegraphics[scale=0.6]{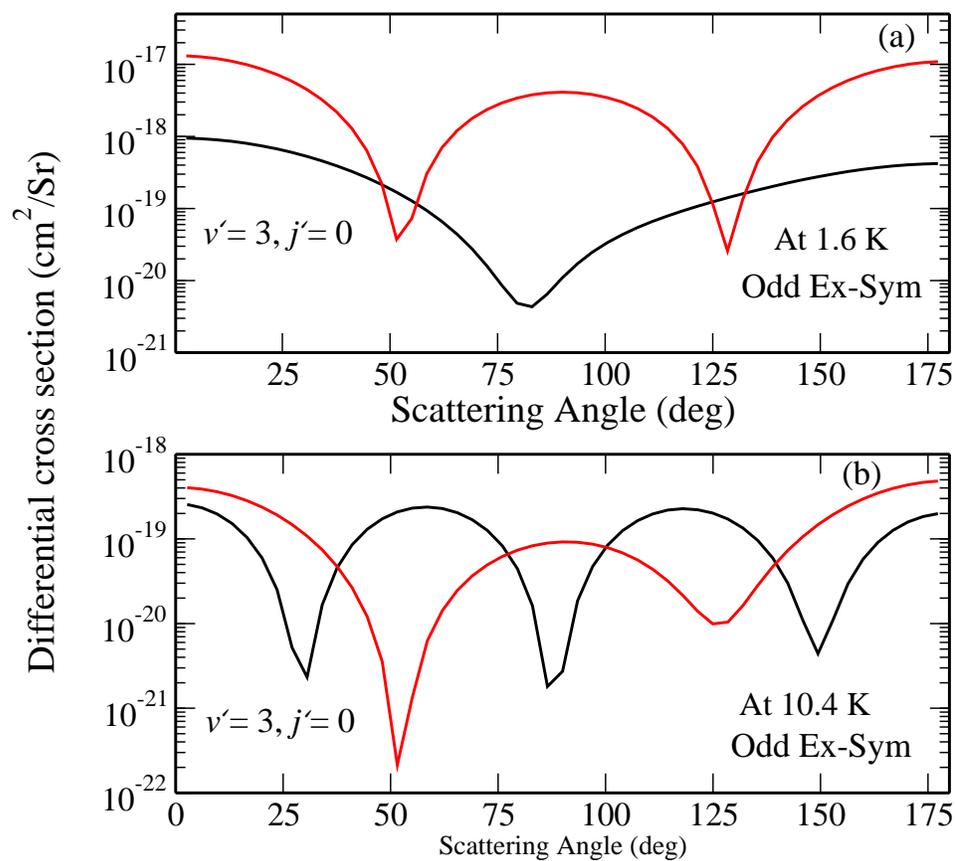}
\caption{
\begin{small} {The DCS for the D + HD($v=4$, $j=0$) $\to$ D + HD($v'=3$, $j'=0$) reaction
(odd exchange symmetry) is plotted as a function of scattering angle for two fixed collision energies:
(a) $E_c = 1.6\,{\rm K}$ and (b) $E_c = 10.4\,{\rm K}$.
The red curves include the geometric phase (GP) and the black curves do not (NGP).
The results include all values of total angular momentum $J=0-4$.}
\end{small}
}
\label{fig2.11}
\end{figure}

\begin{figure}
\includegraphics[scale=0.6]{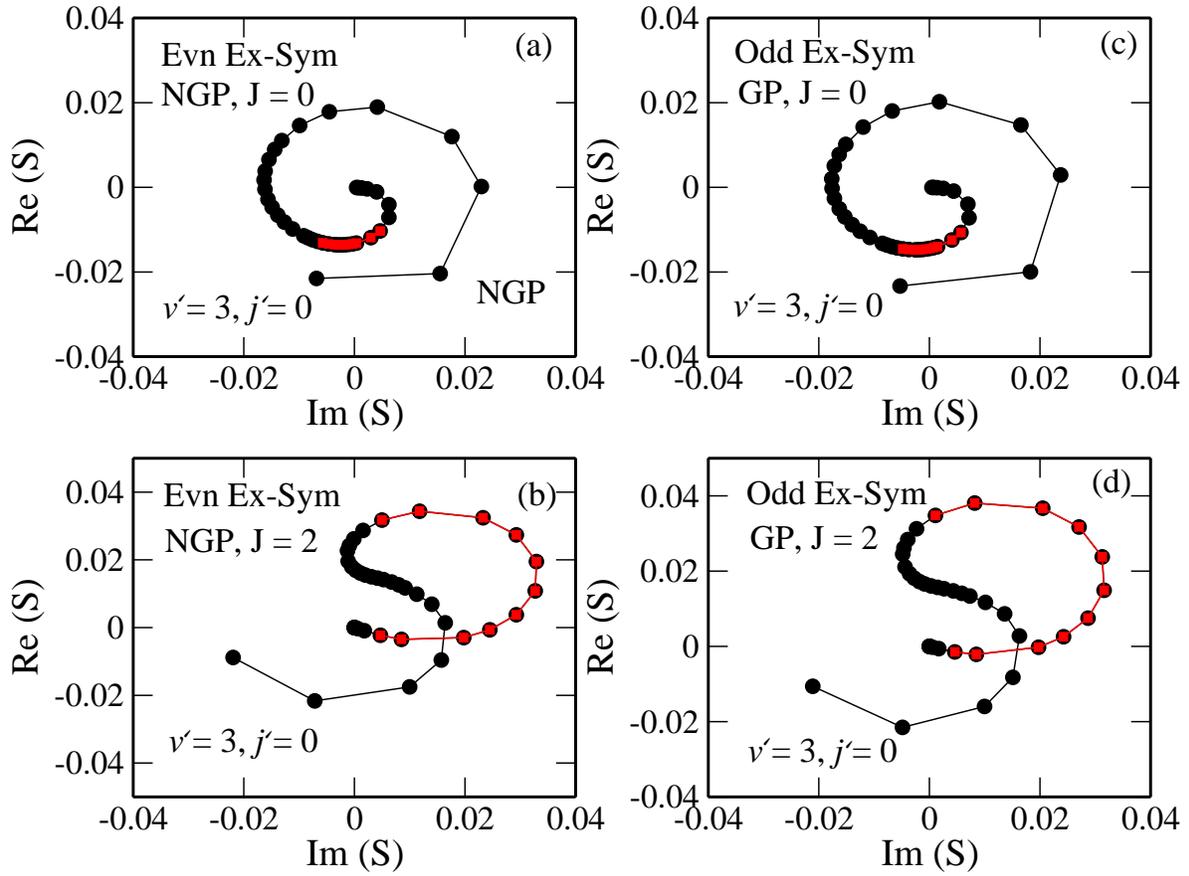}
\caption{
\begin{small} {Argand plots are presented for the D + HD($v=4$, $j=0$) $\to$ D + HD($v'=3$, $j'=0$) reaction
in panels: (a) Even exchange symmetry, $J=0$, and NGP; 
(b) Even exchange symmetry, $J=2$, and NGP; 
(c) Odd exchange symmetry, $J=0$, GP;
(d) Odd exchange symmetry, $J=2$, GP.
The ultracold collision energy at $1\,\mu{\rm K}$ corresponds to the black dot near the origin of the plots.
The red squares indicate collision energies near the $l=2$ shape resonance occuring at $E_c=1.6\,{\rm K}$.}
\end{small}
}
\label{fig2.12}
\end{figure}


\begin{figure}
\includegraphics[scale=0.6]{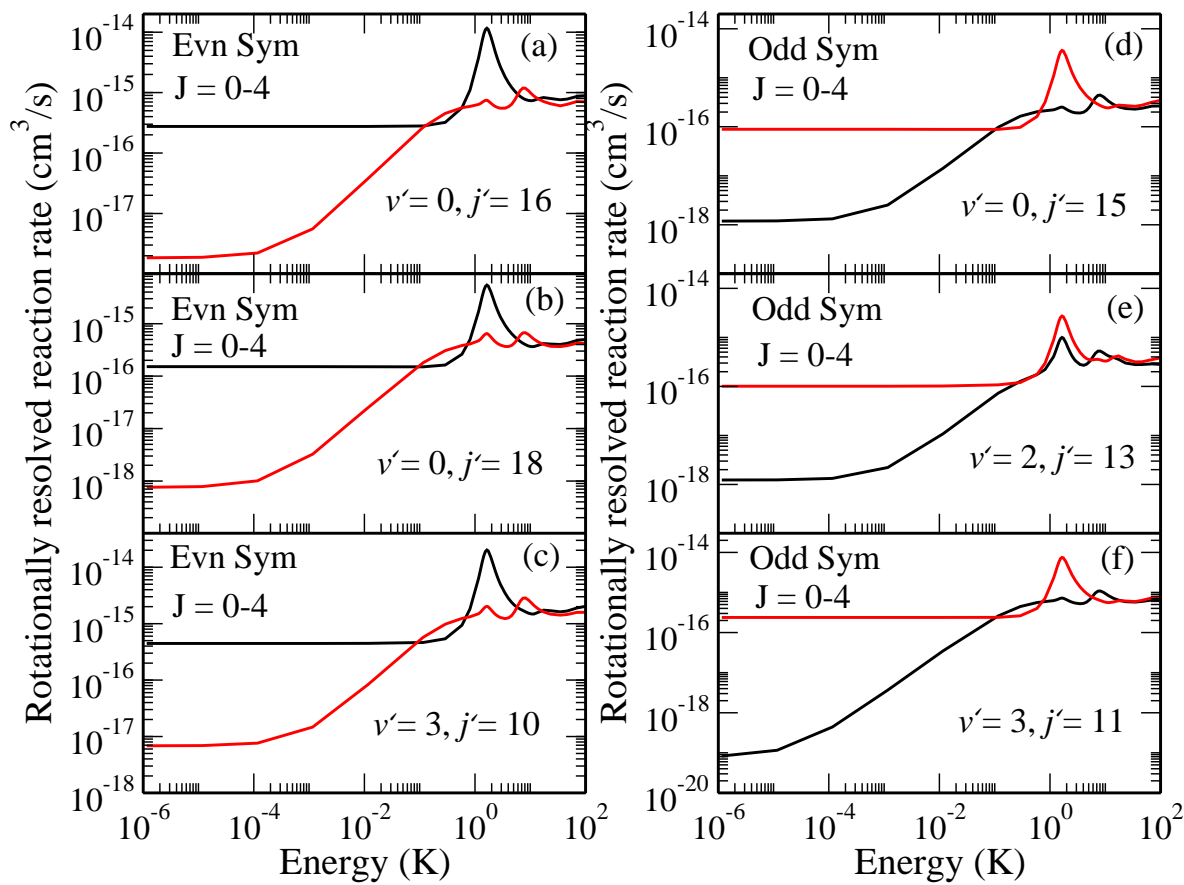}
\caption{
\begin{small} {Rotationally resolved rate coefficients for the D + HD($v=4$, $j=0$) $\to$ H + D$_2$($v'$, $j'$) reaction are plotted
as a function of collision energy:  (a) $v'=0$, $j'=16$, (b) $v'=0$, $j'=18$, (c) $v'=3$, $j'=10$, (d) $v'=0$, $j'=15$, (e) $v'=2$, $j'=13$, 
and (f) $v'=3$, $j'=11$. 
The results for even exchange symmetry are plotted in the left panels (a), (b) and (c), and those for odd exchange symmetry are plotted in
the right panels (d), (e), and (f).  In all panels the red curves include the geometric phase (GP) and the black curves do not (NGP).  
The rates include all values of total angular momentum $J=0-4$.}
\end{small}
}
\label{fig3.1}
\end{figure}

\begin{figure}
\includegraphics[scale=0.6]{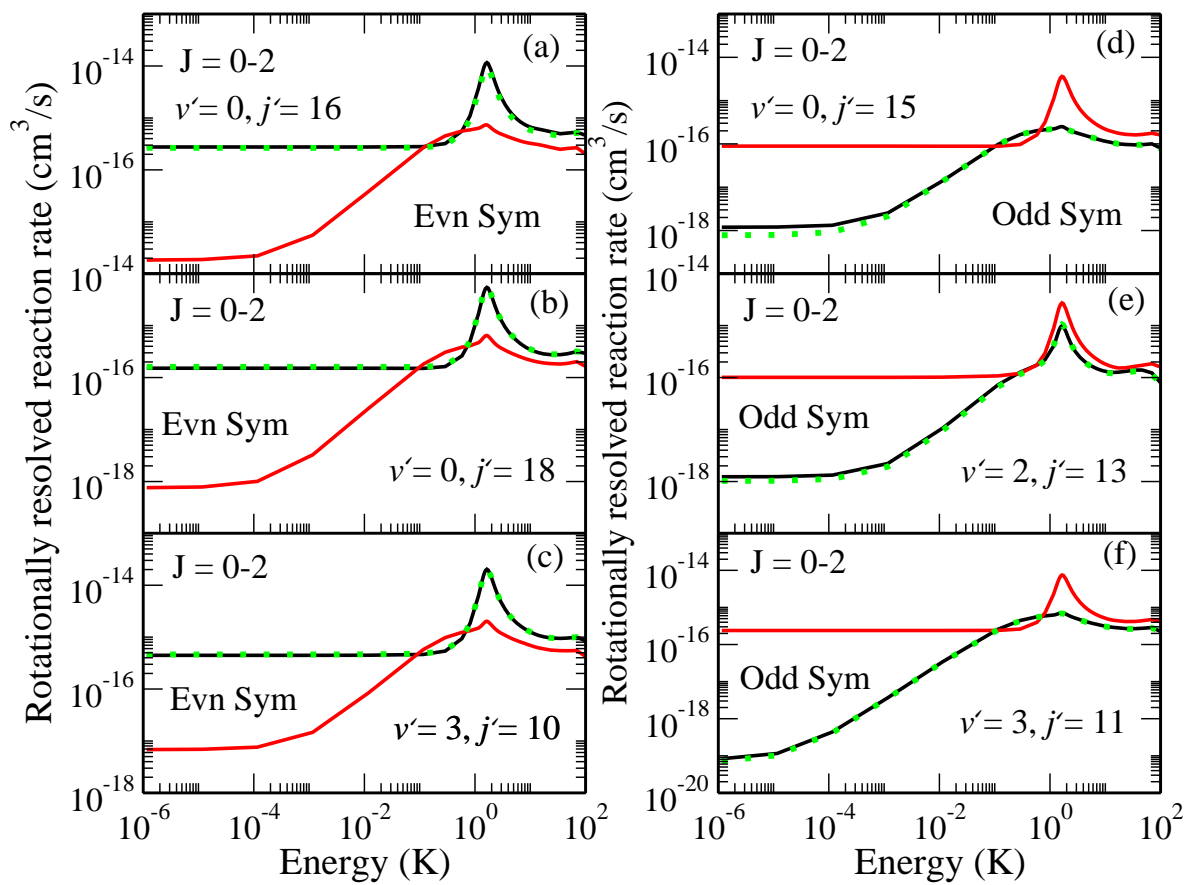}
\caption{
\begin{small} {Gauge invariance check (green squares) for the same product states plotted in Fig. \ref{fig3.1}. Here
only the total values of angular momentum between $J=0-2$ are included.}
\end{small}
}
\label{fig3.3}
\end{figure}

\begin{figure}
\includegraphics[scale=0.6]{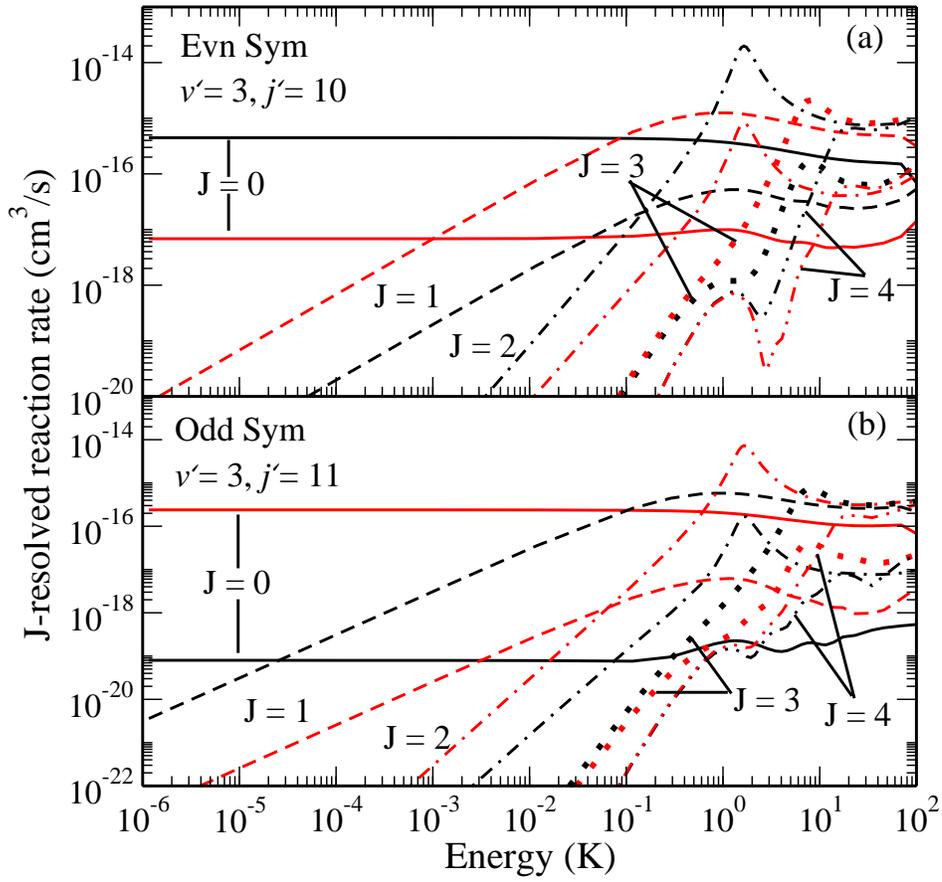}
\caption{
\begin{small} {The $J=l$ resolved rate coefficients for the D + HD($v=4$, $j=0$) $\to$ H + D$_2$($v'=3$, $j'$) reaction are plotted
as a function of collision energy:  (a) $j'=10$ even exchange symmetry and (b) $j'=11$ odd exchange symmetry.
Solid curves $J=0$, dashed curves $J=1$, dot dashed $J=2$, dotted $J=3$, and double-dot dashed $J=4$.
In all cases the black and red curves correspond to NGP and GP, respectively.}
\end{small}
}
\label{fig3.4}
\end{figure}


\begin{figure}
\includegraphics[scale=0.6]{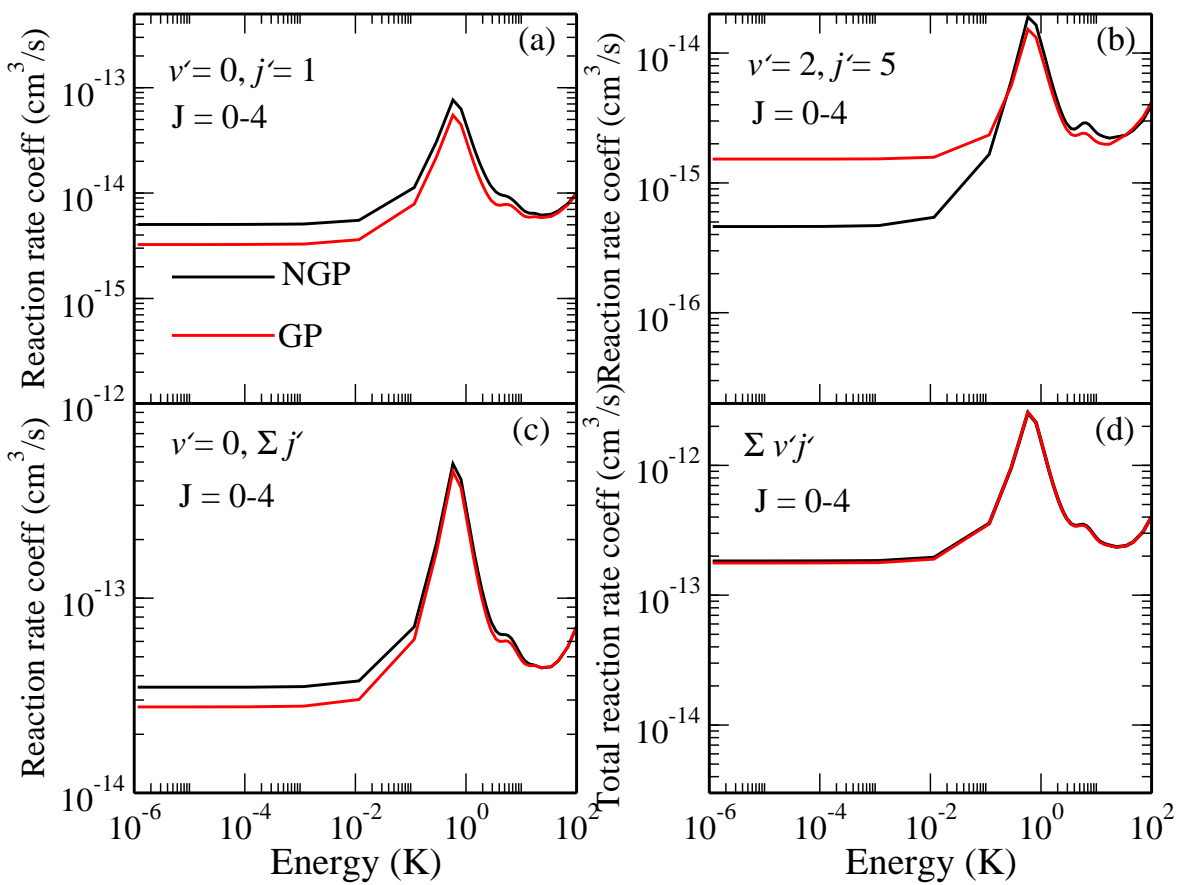}
\caption{Rotationally resolved, vibrationally resolved, and total rate coefficients for the H + D$_2$($v=4$, $j=0$) $\to$ D + HD($v'$, $j'$) reaction are plotted
as a function of collision energy:  (a) $v'=0$, $j'=1$, (b) $v'=2$, $j'=5$, (c) $v'=0$, and (d) total rate.
All of the rates are for even exchange symmetry and include all values of total angular momentum $J=0-4$.
In all cases the black and red curves correspond to NGP and GP, respectively.}
\label{fig4.1}
\end{figure}

\begin{figure}
\includegraphics[scale=0.6]{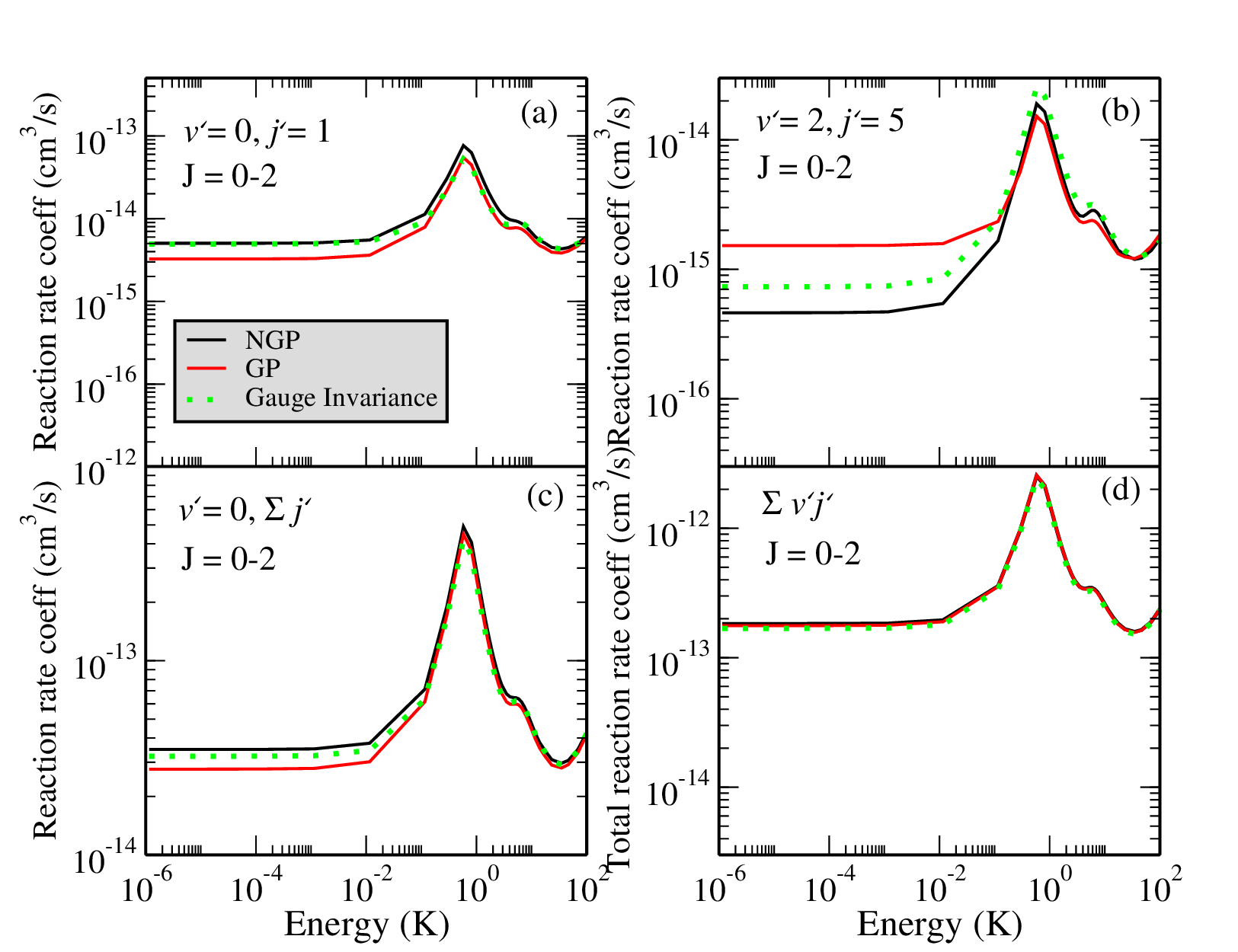}
\caption{
\begin{small} {Gauge invariance check (green squares) for the same product states plotted in Fig. \ref{fig4.1}. Here
only the total values of angular momentum between $J=0-2$ are included.
}
\end{small}
}
\label{fig4.2}
\end{figure}

\begin{table}
\caption{Ultracold ($1\,\mu{\rm K}$) reaction rate coefficients for D + HD($v=4$, $j=0$) $\to$ D + HD($v'$, $j'$) (even exchange symmetry) with and without geometric phase effects, the ratio of the average square modulus of the exchange (reactive) and inelastic (non-reactive) pathways and the corresponding $\cos\Delta$ are tabulated. The rates include the appropriate $2/3$ nuclear spin statistical factor.}
\begin{tabular}{cccccc}
\hline
\hline
$v'$& $j'$ & NGP rate (cm$^3$/s) & GP rate (cm$^3$/s) & $\left<\frac{|f^{\rm ex}|^2}{|f^{\rm inel}|^2}\right>$ & $\cos(\Delta)$\\
\hline
\hline
0  &   0   &  6.57$\times10^{-16}$ & 2.56$\times10^{-17}$  & 0.47  & 0.99 \\
0  &   1   &  5.13$\times10^{-16}$ & 1.35$\times10^{-16}$  & 2.13  & 0.62 \\ 
0  &   2   &  1.43$\times10^{-15}$ & 2.10$\times10^{-16}$  & 0.41  & 0.82 \\ 
0  &   6   &  7.13$\times10^{-17}$ & 1.60$\times10^{-16}$  & 0.71  & -0.93 \\ 
0  &   9   &  4.64$\times10^{-16}$ & 1.50$\times10^{-17}$  & 1.50  & 0.97 \\ 
0   &  11  & 1.11$\times10^{-14}$  & 1.14$\times10^{-18}$  & 1.01  & 0.99 \\  
0   &  12  & 1.13$\times10^{-14}$ & 1.66$\times10^{-18}$  & 0.99  & 0.99 \\ 
0   &  13  & 5.10$\times10^{-15}$ & 3.18$\times10^{-18}$  & 1.09  & 0.99 \\
0   &  14  & 6.64$\times10^{-15}$ & 3.38$\times10^{-18}$  & 1.05  & 0.99 \\
1  &   0   &  7.61$\times10^{-16}$ & 5.80$\times10^{-18}$  & 1.34  & 0.99 \\
1  &   1   &  1.47$\times10^{-15}$ & 5.59$\times10^{-17}$  & 0.76  & 0.93 \\
1  &   2   &  2.41$\times10^{-15}$ & 1.45$\times10^{-16}$  & 0.55  & 0.93 \\
1   &  7    & 5.45$\times10^{-15}$ & 5.60$\times10^{-17}$  & 0.69  & 0.99 \\
1   &  10  & 8.66$\times10^{-15}$ & 1.23$\times10^{-17}$  & 0.97  & 0.99 \\
1   &  11  & 1.64$\times10^{-14}$ & 1.59$\times10^{-17}$  & 1.09  & 0.99 \\
1   &  12  & 1.11$\times10^{-14}$ & 4.68$\times10^{-17}$  & 1.07  & 0.99 \\
2  &   0   &  2.36$\times10^{-15}$ & 2.91$\times10^{-18}$  & 1.01  & 0.99 \\
2  &   1   &  2.18$\times10^{-15}$ & 3.25$\times10^{-17}$  & 0.91  & 0.97 \\
2  &   4   &  1.19$\times10^{-14}$ & 4.95$\times10^{-17}$  & 0.98  & 0.99 \\
2   &  5   & 3.15$\times10^{-14}$ & 2.47$\times10^{-17}$  &  0.94  & 0.98 \\
2   &  6   & 2.08$\times10^{-14}$ & 1.21$\times10^{-16}$  &  1.36  & 0.99 \\
2   &  8   & 2.11$\times10^{-14}$ & 1.30$\times10^{-16}$  &  1.05  & 0.98 \\
2   &  9   & 2.84$\times10^{-14}$ & 1.75$\times10^{-16}$  &  1.27  & 0.99 \\
3  &   0   &  7.93$\times10^{-15}$ & 1.57$\times10^{-16}$  & 1.71  & 0.99 \\
3  &   1   &  2.14$\times10^{-14}$ & 9.46$\times10^{-16}$  & 0.86  & 0.92 \\
3  &   2   &  1.36$\times10^{-14}$ & 4.39$\times10^{-16}$  & 2.05  & 0.99 \\
3   &  3   & 1.99$\times10^{-14}$ & 8.68$\times10^{-16}$  &  1.03  & 0.92 \\
3   &  4   & 3.65$\times10^{-14}$ & 8.55$\times10^{-16}$  &  1.43  & 0.97 \\
3   &  5   & 6.40$\times10^{-14}$ & 6.88$\times10^{-16}$  &  1.38  & 0.99 \\
3   &  6   & 4.51$\times10^{-14}$ & 1.29$\times10^{-15}$  &  1.80  & 0.98 \\
\hline
\hline
\\
\\
\\
\end{tabular}
\label{table1}
\end{table}


\begin{table}
\caption{Ultracold ($1\,\mu{\rm K}$) reaction rate coefficients for D + HD($v=4$, $j=0$) $\to$ D + HD($v'$, $j'$) (odd exchange symmetry) with and without geometric phase effects, the ratio of the average square modulus of the exchange (reactive) and inelastic (non-reactive) pathways and the corresponding $\cos\Delta$ are tabulated. The rates include the appropriate $1/3$ nuclear spin statistical factor.}
\begin{tabular}{cccccc}
\hline
\hline
$v'$& $j'$ & NGP rate (cm$^3$/s) & GP rate (cm$^3$/s) & $\left<\frac{|f^{\rm ex}|^2}{|f^{\rm inel}|^2}\right>$ & $\cos(\Delta)$\\
\hline
\hline
0  &   0   &  1.13$\times10^{-17}$ & 3.34$\times10^{-16}$  & 0.49  & -0.99\\
0  &   1   &  6.12$\times10^{-17}$ & 2.69$\times10^{-16}$  & 1.93  & -0.66 \\ 
0  &   2   &  9.01$\times10^{-17}$ & 7.10$\times10^{-16}$  & 0.47  & -0.83 \\ 
0  &   6   &  8.06$\times10^{-17}$ & 2.16$\times10^{-18}$  & 0.72  & 0.96 \\ 
0  &   9   &  1.03$\times10^{-17}$ & 2.37$\times10^{-16}$  & 1.41  & -0.93 \\ 
0   &  11  & 6.29$\times10^{-19}$  & 5.46$\times10^{-15}$ & 1.03  & -0.99 \\  
0   &  12  & 1.05$\times10^{-18}$  & 5.52$\times10^{-15}$ & 0.97  & -0.99 \\ 
0   &  13  & 2.01$\times10^{-18}$  & 2.55$\times10^{-15}$ & 1.03  & -0.99 \\ 
0   &  14  & 2.86$\times10^{-18}$  & 3.37$\times10^{-15}$ & 1.02  & -0.99 \\ 
1  &   0   &  2.06$\times10^{-18}$ & 3.80$\times10^{-16}$  & 1.04  & -0.99 \\
1  &   1   &  2.92$\times10^{-17}$ & 7.09$\times10^{-16}$  & 0.71  & -0.93 \\
1  &   2   &  6.30$\times10^{-17}$ & 1.17$\times10^{-15}$  & 0.56  & -0.93 \\
1   &  7    & 3.66$\times10^{-17}$  & 2.62$\times10^{-15}$  & 0.65  & -0.99 \\
1   &  10  & 7.04$\times10^{-18}$  & 4.24$\times10^{-15}$  & 0.95  & -0.99 \\
1   &  11  & 9.51$\times10^{-18}$  & 8.39$\times10^{-15}$  & 1.09  & -0.99 \\
1   &  12  & 2.18$\times10^{-18}$  & 5.57$\times10^{-15}$  & 1.19  & -0.99 \\
2  &   0   &  6.38$\times10^{-18}$ & 1.17$\times10^{-15}$  & 0.94  & -0.99 \\
2  &   1   &  2.40$\times10^{-17}$ & 1.10$\times10^{-15}$  & 0.93  & -0.96 \\
2  &   4   &  3.36$\times10^{-17}$ & 5.92$\times10^{-15}$  & 0.99  & -0.99 \\
2   &  5   &  1.33$\times10^{-16}$ & 1.56$\times10^{-14}$  &  0.94  & -0.98 \\
2   &  6   &  5.63$\times10^{-17}$ & 1.03$\times10^{-14}$  &  1.34  & -0.99 \\
2   &  8   &  6.54$\times10^{-17}$ & 1.05$\times10^{-14}$  &  1.02  & -0.98 \\
2   &  9   & 8.17$\times10^{-17}$ & 1.40$\times10^{-14}$   &  1.27  & -0.99 \\
3  &   0   & 9.49$\times10^{-17}$ & 4.58$\times10^{-15}$  & 1.78  & -0.99 \\
3  &   1   & 5.03$\times10^{-16}$ & 1.07$\times10^{-14}$  & 0.87  & -0.91 \\
3  &   2   & 2.21$\times10^{-16}$ & 6.75$\times10^{-15}$  & 2.06  & -0.99 \\
3   &  3   & 4.44$\times10^{-16}$ & 9.97$\times10^{-15}$  &   0.99  & -0.91 \\
3   &  4   & 4.05$\times10^{-16}$ & 1.84$\times10^{-14}$  &  1.42  & -0.97 \\
3   &  5   & 3.66$\times10^{-16}$ & 3.24$\times10^{-14}$  &  1.40  & -0.99 \\
3   &  6   & 6.18$\times10^{-16}$ & 2.26$\times10^{-14}$  &  1.78  & -0.98 \\
\hline
\hline
\end{tabular}
\label{table2}
\end{table}

\end{document}